\documentclass[fleqn,usenatbib]{mnras}

\usepackage{newtxtext,newtxmath}

\usepackage[T1]{fontenc}
\usepackage{ae,aecompl}

\usepackage{graphicx}	

\newcommand{\todo}{\ifmmode \text{\color{red}\Huge{\(\bullet\)}} \else {\color{red}{\Huge$\bullet$}}\fi}
\newcommand{\tido}{\ifmmode {{\color{red}\bullet}} \else {\color{red}$\bullet$}\fi}

\newcommand{\E        }[1]{\ifmmode 10^{#1} \else $10^{#1}$\fi}
\newcommand{\tE        }[1]{\ifmmode \times10^{#1} \else $\times10^{#1}$\fi}
\newcommand{\til}{\ifmmode \sim \else $\sim$\fi}
\renewcommand{\~} {\ifmmode \sim \else $\sim$\fi}

\newcommand{\et}{et al.\ }

\newcommand{\pc}	{\ifmmode {\rm pc} \else pc\fi}
\newcommand{\kpc}	{\ifmmode {\rm kpc} \else kpc\fi}
\newcommand{\ld}	{\ifmmode {\rm l.d.} \else l.d.\fi}
\newcommand{\kms}	{\ifmmode {\rm km\,s}^{-1} \else km\,s$^{-1}$\fi}
\newcommand{\cc}	{\ifmmode {\rm cm}^{-3}    \else cm$^{-3}$\fi}
\newcommand{\cmii}	{\ifmmode {\rm cm}^{-2}    \else cm$^{-2}$\fi}
\newcommand{\ergs}	{\ifmmode {\rm erg\,s}^{-1} \else erg s$^{-1}$\fi}
\newcommand{\ergcms}	{\ifmmode {\rm erg\,cm}^{-2}\,{\rm s}^{-1} \else erg\,cm$^{-2}$\,s$^{-1}$\fi}
\newcommand{\ergcmsA}	{\ifmmode {\rm erg\,cm}^{-2}\,{\rm s}^{-1}\,{\rm\AA}^{-1}
\else erg\,cm$^{-2}$\,s$^{-1}$\,\AA$^{-1}$\fi}
\newcommand{  \ergcmsHz  }{\ifmmode{\rm erg\,cm}^{-2}\,{\rm s}^{-1}\,{\rm Hz}^{-1}
                       \else ergs\,cm$^{-2}$\,s$^{-1}$\,Hz$^{-1}$\fi}
\newcommand{\kev}	{\ifmmode {\rm keV} \else keV\fi}

\newcommand{\mic}	{\ifmmode {\rm \mu m} \else $\mu$m\fi}
\newcommand{\vFWHM}	{\ifmmode v_{\mbox{\tiny FWHM}} \else $v_{\mbox{\tiny FWHM}}$\fi}
\newcommand{\vBLR}	{\ifmmode v_{\mbox{\tiny BLR}} \else $v_{\mbox{\tiny BLR}}$\fi}
\newcommand{\sigBLR}	{\ifmmode \sigma_{\mbox{\tiny BLR}} \else $\sigma_{\mbox{\tiny BLR}}$\fi}
\newcommand{\vNLR}	{\ifmmode v_{\mbox{\tiny NLR}} \else $v_{\mbox{\tiny NLR}}$\fi}
\newcommand{\tauBLR}	{\ifmmode \tau_{\mbox{\tiny BLR}} \else $\tau_{\mbox{\tiny BLR}}$\fi}

\newcommand{\Hubble}	{\ifmmode {\rm km\,s}^{-1}\,{\rm Mpc}^{-1} \else km\,s$^{-1}$\,Mpc$^{-1}$\fi}
\newcommand{\NDunit}	{\ifmmode {\rm Mpc}^{-3} \else Mpc$^{-3}$\fi}
\newcommand{\LFunit}	{\ifmmode {\rm Mpc}^{-3}\,{\rm mag}^{-1} \else Mpc$^{-3}$\,mag$^{-1}$\fi}
\newcommand{\MFunit}	{\ifmmode {\rm Mpc}^{-3}\,{\rm dex}^{-1} \else Mpc$^{-3}$\,dex$^{-1}$\fi}

\newcommand{\Msun}{\ifmmode M_{\odot} \else $M_{\odot}$\fi}
\newcommand{\Lsun}{\ifmmode L_{\odot} \else $L_{\odot}$\fi}
\newcommand{\Zsun}{\ifmmode Z_{\odot} \else $Z_{\odot}$\fi}
\newcommand{\mpyr}{\ifmmode \Msun\,{\rm yr}^{-1} \else $\Msun\,{\rm yr}^{-1}$\fi}

\newcommand{\Msol}{\Msun}

\newcommand{\qnote}{\ifmmode q_{0} \else $q_{0}$\fi}
\newcommand{\Hnote}{\ifmmode H_{0} \else $H_{0}$\fi}
\newcommand{\hnote}{\ifmmode h_{0} \else $h_{0}$\fi}
\newcommand{\anote}{\ifmmode a_{0} \else $a_{0}$\fi}
\newcommand{\tnote}{\ifmmode t_{0} \else $t_{0}$\fi}

\newcommand{\ltsim}{\raisebox{-.5ex}{$\;\stackrel{<}{\sim}\;$}}

\def\gsim{\;\rlap{\lower 2.5pt \hbox{$\sim$}}\raise 1.5pt\hbox{$>$}\;}
\def\lsim{\;\rlap{\lower 2.5pt \hbox{$\sim$}}\raise 1.5pt\hbox{$<$}\;}

\newcommand{  \Halpha   }{\ifmmode {\rm H}\alpha \else H$\alpha$\fi}
\newcommand{  \halpha   }{\Halpha}
\newcommand{  \ha       }{\Halpha}
\newcommand{  \Hbeta    }{\ifmmode {\rm H}\beta \else H$\beta$\fi}
\newcommand{  \hbeta    }{\Hbeta}
\newcommand{  \hb       }{\Hbeta}
\newcommand{  \Hgamma   }{\ifmmode {\rm H}\gamma \else H$\gamma$\fi}
\newcommand{  \Hdelta   }{\ifmmode {\rm H}\delta \else H$\delta$\fi}
\newcommand{  \Lya      }{\ifmmode {\rm Ly}\alpha \else Ly$\alpha$\fi}
\newcommand{  \Lyb      }{\ifmmode {\rm Ly}\beta \else Ly$\beta$\fi}
\newcommand{  \Pa       }{\ifmmode {\rm P}\alpha \else P$\alpha$\fi}
\newcommand{  \Pb       }{\ifmmode {\rm P}\beta \else P$\beta$\fi}
\newcommand{  \Bra      }{\ifmmode {\rm Br}\alpha \else Br$\alpha$\fi}
\newcommand{  \Brg      }{\ifmmode {\rm Br}\gamma \else Br$\gamma$\fi}
\newcommand{  \hii      }{\ifmmode {\rm H}\,\textsc{ii} \else H\,\textsc{ii}\fi}
\newcommand{  \hei      }{\ifmmode {\rm He}\,\textsc{i} \else He\,\textsc{i}\fi}
\newcommand{  \heii     }{\ifmmode {\rm He}\,\textsc{ii} \else He\,\textsc{ii}\fi}
\newcommand{  \HeIIuv   }{\ifmmode {\rm He}\,\textsc{ii}\,\lambda1640 \else He\,\textsc{ii}\,$\lambda1640$\fi}
\newcommand{  \HeIIop   }{\ifmmode {\rm He}\,\textsc{ii}\,\lambda4686 \else He\,\textsc{ii}\,$\lambda4686$\fi}
\newcommand{  \CII	}{\ifmmode \left[{\rm C}\,\textsc{ii}\right]\,\lambda157.74\,\mu{\rm m} \else [C\,{\sc ii}]\ $\lambda157.74\,\mu{\rm m}$\fi}
\newcommand{  \cii	}{\ifmmode \left[{\rm C}\,\textsc{ii}\right] \else [C\,{\sc ii}]\fi}

\newcommand{  \ciii     }{\ifmmode {\rm C}\,\textsc{iii}\right] \else C\,\textsc{iii}]\fi}
\newcommand{  \CIII     }{\ifmmode {\rm C}\,\textsc{iii}\right]\,\lambda1909 \else C\,\textsc{iii}]\,$\lambda1909$\fi}
\newcommand{  \civ      }{\ifmmode {\rm C}\,\textsc{iv}  \else C\,\textsc{iv}\fi}
\newcommand{  \CIV      }{\ifmmode {\rm C}\,\textsc{iv}\,\lambda1549 \else C\,\textsc{iv}\,$\lambda1549$\fi}
\newcommand{  \nii      }{\ifmmode {\rm N}\,\textsc{ii}  \else N\,\textsc{ii}\fi}
\newcommand{  \niii     }{\ifmmode {\rm N}\,\textsc{iii} \else N\,\textsc{iii}\fi}
\newcommand{  \niv      }{\ifmmode {\rm N}\,\textsc{iv}  \else N\,\textsc{iv}\fi}
\newcommand{  \NIVuv    }{\ifmmode {\rm N}\,\textsc{iv}\,\lambda1486 \else N\,\textsc{iv}\,$\lambda1486$\fi}
\newcommand{  \nv       }{\ifmmode {\rm N}\,\textsc{v}   \else N\,\textsc{v}\fi}
\newcommand{\oi}{\ifmmode \left[{\rm O}\,\textsc{i}\right] \else [O\,{\sc i}]\fi}
\newcommand{\OI}{\ifmmode \left[{\rm O}\,\textsc{i}\right]\,\lambda6300 \else [O\,{\sc i}]$\,\lambda6300$\fi}
\newcommand{\oii}{\ifmmode \left[{\rm O}\,\textsc{ii}\right] \else [O\,{\sc ii}]\fi}
\newcommand{\OII}{\ifmmode \left[{\rm O}\,\textsc{ii}\right]\,\lambda3727 \else [O\,{\sc ii}]\,$\lambda3727$\fi}
\newcommand{\oiii}{\ifmmode \left[{\rm O}\,\textsc{iii}\right] \else [O\,{\sc iii}]\fi}
\newcommand{\OIII}{\ifmmode \left[{\rm O}\,\textsc{iii}\right]\,\lambda5007 \else [O\,{\sc iii}]\,$\lambda5007$\fi}
\newcommand{  \OIIIuv   }{\ifmmode {\rm O}\,\textsc{iii}\,\lambda1663 \else O\,\textsc{iii}\,$\lambda1663$\fi}
\newcommand{  \oiv      }{\ifmmode {\rm O}\,\textsc{iv}  \else O\,\textsc{iv}\fi}
\newcommand{  \OIVuv    }{\ifmmode {\rm O}\,\textsc{iv}\,\lambda1402  \else O\,\textsc{iv}\,$\lambda1402$\fi}
\newcommand{  \OIVIR    }{\ifmmode {\rm O}\,\textsc{iv}\,25.9\,\mu {\rm m} \else O\,\textsc{iv}\,$25.9\,\mu$m\fi}
\newcommand{  \ovi      }{\ifmmode {\rm O}\,\textsc{vi}   \else O\,\textsc{vi}\fi}
\newcommand{  \Ovi      }{\ifmmode {\rm O}\,\textsc{vi}\,\lambda1035 \else O\,\textsc{vi}\,$\lambda1035$\fi}
\newcommand{  \nei      }{\ifmmode {\rm Ne}\,\textsc{i}   \else Ne\,\textsc{i}\fi}
\newcommand{  \neii     }{\ifmmode {\rm Ne}\,\textsc{ii}  \else Ne\,\textsc{ii}\fi}
\newcommand{  \NeiiIR   }{\ifmmode {\rm Ne}\,\textsc{ii}\,12.8\,\mu {\rm m} \else Ne\,\textsc{ii}\,$12.8\,\mu$m\fi}
\newcommand{  \neiii    }{\ifmmode {\rm Ne}\,\textsc{iii} \else Ne\,\textsc{iii}\fi}
\newcommand{  \neiv     }{\ifmmode {\rm Ne}\,\textsc{iv}  \else Ne\,\textsc{iv}\fi}
\newcommand{  \nev      }{\ifmmode {\rm Ne}\,\textsc{v}   \else Ne\,\textsc{v}\fi}
\newcommand{  \NevIR    }{\ifmmode {\rm Ne}\,\textsc{v}\,24.3\,\mu {\rm m} \else Ne\,\textsc{v}\,$24.3\,\mu$m\fi}
\newcommand{  \nevi     }{\ifmmode {\rm Ne}\,\textsc{vi}  \else Ne\,\textsc{vi}\fi}
\newcommand{  \mgi      }{\ifmmode {\rm Mg}\,\textsc{i} \else Mg\,\textsc{i}\fi}
\newcommand{  \mgii     }{\ifmmode {\rm Mg}\,\textsc{ii} \else Mg\,\textsc{ii}\fi}
\newcommand{  \MgII     }{\ifmmode {\rm Mg}\,\textsc{ii}\,\lambda2798 \else Mg\,\textsc{ii}\,$\lambda2798$\fi}
\newcommand{  \sii      }{\ifmmode {\rm S}\,\textsc{ii} \else S\,\textsc{ii}\fi}
\newcommand{  \siii     }{\ifmmode {\rm S}\,\textsc{iii} \else S\,\textsc{iii}\fi}
\newcommand{  \siv      }{\ifmmode {\rm S}\,\textsc{iv} \else S\,\textsc{iv}\fi}
\newcommand{  \sili     }{\ifmmode {\rm Si}\,\textsc{i}   \else Si\,\textsc{i}\fi}
\newcommand{  \silii    }{\ifmmode {\rm Si}\,\textsc{ii}  \else Si\,\textsc{ii}\fi}
\newcommand{  \Siliv    }{\ifmmode {\rm Si}\,\textsc{iv}  \else Si\,\textsc{iv}\fi}
\newcommand{  \SilIVuv  }{\ifmmode {\rm Si}\,\textsc{iv}\,\lambda1400  \else Si\,\textsc{iv}\,$\lambda1400$\fi}
\newcommand{  \AlIII   }{\ifmmode {\rm Al}\,\textsc{iii}\,\lambda1857 \else Al\,\textsc{iii}\,$\lambda1857$\fi}
\newcommand{  \Aliii   }{\ifmmode {\rm Al}\,\textsc{iii} \else Al\,\textsc{iii}\fi}
\newcommand{  \caii     }{\ifmmode {\rm Ca}\,\textsc{ii} \else Ca\,\textsc{ii}\fi}
\newcommand{  \feii     }{\ifmmode {\rm Fe}\,\textsc{ii} \else Fe\,\textsc{ii}\fi}
\newcommand{  \feiii    }{\ifmmode {\rm Fe}\,\textsc{iii} \else Fe\,\textsc{iii}\fi}
\newcommand{  \Kalpha   }{\ifmmode {\rm K}\alpha \else K$\alpha$\fi}

\newcommand{ \Lhb   }{\ifmmode L_{\hb} \else $L_{\hb}$\fi}
\newcommand{ \Lha   }{\ifmmode L_{\ha} \else $L_{\ha}$\fi}
\newcommand{ \fwhb  }{\ifmmode {\rm FWHM}\left(\hb\right) \else FWHM(\hb)\fi}
\newcommand{\sighb  }{\ifmmode \sigma\left(\hb\right) \else $\sigma\left(\hb\right)$\fi}
\newcommand{ \ewhb  }{\ifmmode {\rm EW}\left(\hb\right) \else EW(\hb)\fi}
\newcommand{ \fwha  }{\ifmmode {\rm FWHM}\left(\ha\right) \else FWHM(\ha)\fi}
\newcommand{ \ewha  }{\ifmmode {\rm EW}\left(\ha\right) \else EW(\ha)\fi}
\newcommand{ \Lmg   }{\ifmmode L\left(\mgii\right) \else $L\left(\mgii\right)$\fi}
\newcommand{ \fwmg  }{\ifmmode {\rm FWHM}\left(\mgii\right) \else FWHM(\mgii)\fi}
\newcommand{ \Lciv  }{\ifmmode L\left(\civ\right) \else $L\left(\civ\right)$\fi}
\newcommand{ \fwciv }{\ifmmode {\rm FWHM}\left(\civ\right) \else FWHM(\civ)\fi}
\newcommand{ \fwhm  }{\ifmmode {\rm FWHM} \else FWHM\fi} 
\newcommand{ \voff  }{\ifmmode v_{\rm off} \else $v_{\rm off}$\fi} 
\newcommand{ \vmax  }{\ifmmode v_{\rm max} \else $v_{\rm max}$\fi} 

\newcommand{ \mumg  }{\ifmmode \mu\left(\mgii\right) \else $\mu\left(\mgii\right)$\fi}
\newcommand{ \fmg   }{\ifmmode f\left(\mgii\right) \else $f\left(\mgii\right)$\fi}
\newcommand{ \muciv }{\ifmmode \mu\left(\civ\right) \else $\mu\left(\civ\right)$\fi}
\newcommand{ \fciv  }{\ifmmode f\left(\civ\right) \else $f\left(\civ\right)$\fi}

\newcommand{  \auvo     }{\ifmmode \alpha_{\nu,{\rm UVO}} \else $\alpha_{\nu,{\rm UVO}}$\fi}
\newcommand{  \Ledd     }{\ifmmode L_{\rm Edd} \else $L_{\rm Edd}$\fi}
\newcommand{  \lamLlam  }{\ifmmode \lambda {L_\lambda} \else $\lambda {L_\lambda}$\fi}
\newcommand{  \lLl      }{\ifmmode \lambda {L_\lambda} \else $\lambda {L_\lambda}$\fi}
\newcommand{  \nuLnu    }{\ifmmode \nu L_{\nu} \else $\nu L_{\nu}$\fi}
\newcommand{  \nLn      }{\ifmmode \nu L_{\nu} \else $\nu L_{\nu}$\fi}
\newcommand{  \Luv      }{\ifmmode L_{1450} \else $L_{1450}$\fi}
\newcommand{  \Lop      }{\ifmmode L_{5100} \else $L_{5100}$\fi}
\newcommand{  \lLop     }{\ifmmode \log\left(\Lop/\ergs\right) \else $\log\left(\Lop/\ergs\right)$\fi}
\newcommand{  \Lthree   }{\ifmmode L_{3000} \else $L_{3000}$\fi}
\newcommand{  \lLthree  }{\ifmmode \log\left(\Lthree/\ergs\right) \else $\log\left(\Lthree/\ergs\right)$\fi}
\newcommand{  \Lsix      }{\ifmmode L_{6200} \else $L_{6200}$\fi}
\newcommand{  \lLisx     }{\ifmmode \log\left(\Lop/\ergs\right) \else $\log\left(\Lop/\ergs\right)$\fi}
\newcommand{  \Lxray    }{\ifmmode L_{\rm X} \else $L_{\rm X}$\fi}
\newcommand{  \Lx    }{\Lxray}
\newcommand{  \Lhard    }{\ifmmode L_{\rm 2-10} \else $L_{\rm 2-10}$\fi}
\newcommand{  \Lsoft    }{\ifmmode L_{\rm 0.5-2} \else $L_{\rm 0.5-2}$\fi}

\newcommand{\Fthree}{\ifmmode F_{3000} \else $F_{3000}$\fi}
\newcommand{\fuv}{\ifmmode f_{\lambda}\left(1450{\rm \AA}\right) \else $f_{\lambda}\left(1450 {\rm \AA}\right)$\fi}
\newcommand{\fthree}{\ifmmode f_{\lambda}\left(3000{\rm \AA}\right) \else $f_{\lambda}\left(3000{\rm \AA}\right)$\fi}
\newcommand{\fH}{\ifmmode f_{\lambda}\left(1.65\micron\right) \else
$f_{\lambda}\left(1.65\micron\right)$\fi}

\newcommand{\fbol}{\ifmmode f_{\rm bol} \else $f_{\rm bol}$\fi}
\newcommand{\fbolwv}{\ifmmode f_{\rm bol}\left(\lambda\right) \else $f_{\rm bol}\left(\lambda\right)$\fi}
\newcommand{\fbolopt}{\ifmmode f_{\rm bol}\left(5100{\rm \AA}\right) \else $f_{\rm bol}\left(5100{\rm \AA}\right)$\fi}
\newcommand{\fbolthree}{\ifmmode f_{\rm bol}\left(3000{\rm \AA}\right) \else $f_{\rm bol}\left(3000{\rm \AA}\right)$\fi}
\newcommand{\fboluv}{\ifmmode f_{\rm bol}\left(1450{\rm \AA}\right) \else $f_{\rm bol}\left(1450{\rm \AA}\right)$\fi}

\newcommand{\fbolbat}{\ifmmode f_{\rm bol,\, 14-150\,\kev} \else $f_{\rm bol,\, 14-150\,\kev}$\fi}
\newcommand{\fbolhard}{\ifmmode f_{\rm bol,\, 2-10\,\kev} \else $f_{\rm bol,\, 2-10\,\kev}$\fi}

\newcommand{\fobs}{\ifmmode f_{\rm obs} \else $f_{\rm obs}$\fi}

\newcommand{  \mbh      }{\ifmmode M_{\rm BH} \else $M_{\rm BH}$\fi}
\newcommand{  \lmbh     }{\ifmmode \log\left(\mbh/\Msun\right) \else $\log\left(\mbh/\Msun\right)$\fi} 
\newcommand{  \lledd    }{\ifmmode L/L_{\rm Edd} \else $L/L_{\rm Edd}$\fi}
\newcommand{  \mmedd    }{\ifmmode \dot{m}/\dot{m}_{\rm \,Edd} \else $\dot{m}/\dot{m}_{\rm \,Edd}$\fi}
\newcommand{  \Lbol     }{\ifmmode L_{\rm bol} \else $L_{\rm bol}$\fi}
\newcommand{  \lbol     }{\ifmmode L_{\rm bol} \else $L_{\rm bol}$\fi}
\newcommand{  \lLbol    }{\ifmmode \log\left(\Lbol/\ergs\right) \else $\log\left(\Lbol/\ergs\right)$\fi} 
\newcommand{  \Lagn     }{\ifmmode L_{\rm AGN} \else $L_{\rm AGN}$\fi}
\newcommand{  \lagn     }{\ifmmode L_{\rm AGN} \else $L_{\rm AGN}$\fi}

\newcommand{  \tgrow     }{\ifmmode t_{\rm growth} \else $t_{\rm growth}$\fi}
\newcommand{  \tAD     }{\ifmmode t_{\rm acc} \else $t_{\rm acc}$\fi}
\newcommand{  \tacc    }{\ifmmode t_{\rm acc} \else $t_{\rm acc}$\fi}
\newcommand{  \tUni      }{\ifmmode t_{\rm Universe} \else $t_{\rm Universe}$\fi}

\newcommand{  \Mdotin	}{\ifmmode \dot{M}_{\rm infall} \else $\dot{M}_{\rm infall}$\fi}
\newcommand{  \Mdotbh	}{\ifmmode \dot{M}_{\rm BH} \else $\dot{M}_{\rm BH}$\fi}
\newcommand{  \Mdotad	}{\ifmmode \dot{M}_{\rm AD} \else $\dot{M}_{\rm AD}$\fi}
\newcommand{  \Mdotacc	}{\ifmmode \dot{M}_{\rm acc} \else $\dot{M}_{\rm acc}$\fi}
\newcommand{  \Mdotthin	}{\ifmmode \dot{M}_{\rm thin} \else $\dot{M}_{\rm thin}$\fi}
\newcommand{  \Mdotdisk	}{\ifmmode \dot{M}_{\rm disk} \else $\dot{M}_{\rm disk}$\fi}

\newcommand{  \Mindot	}{\ifmmode \dot{M}_{\rm infall} \else $\dot{M}_{\rm infall}$\fi}
\newcommand{  \Mbhdot	}{\ifmmode \dot{M}_{\rm BH} \else $\dot{M}_{\rm BH}$\fi}
\newcommand{  \Maddot	}{\ifmmode \dot{M}_{\rm AD} \else $\dot{M}_{\rm AD}$\fi}
\newcommand{  \Maccdot	}{\ifmmode \dot{M}_{\rm acc} \else $\dot{M}_{\rm acc}$\fi}
\newcommand{  \Mthdot	}{\ifmmode \dot{M}_{\rm thin} \else $\dot{M}_{\rm thin}$\fi}
\newcommand{  \Mdsdot	}{\ifmmode \dot{M}_{\rm disk} \else $\dot{M}_{\rm disk}$\fi}

\newcommand{  \as	}{\ifmmode a_{\rm *} \else $a_{\rm *}$\fi}
\newcommand{  \avec	}{\ifmmode \vec{a}_{\rm *} \else $\vec{a}_{\rm *}$\fi}
\newcommand{  \re	}{\ifmmode \eta      	 \else $\eta$\fi}
\newcommand{  \RISCO	}{\ifmmode R_{\rm ISCO}  \else $R_{\rm ISCO}$\fi}

\newcommand{  \mseed    }{\ifmmode M_{\rm seed} \else $M_{\rm seed}$\fi}
\newcommand{  \mbul     }{\ifmmode M_{\rm bulge} \else $M_{\rm bulge}$\fi} 
\newcommand{  \mstar    }{\ifmmode M_{*} \else $M_{*}$\fi} 
\newcommand{  \mgal     }{\ifmmode M_{*} \else $M_{*}$\fi} 
\newcommand{  \mhost    }{\ifmmode M_{\rm host} \else $M_{\rm host}$\fi}
\newcommand{  \mmsmall  }{\ifmmode M_{\rm BH}/M_{*} \else $M_{\rm BH}/M_{*}$\fi}
\newcommand{  \mmlarge  }{\ifmmode M_{*}/M_{\rm BH} \else $M_{*}/M_{\rm BH}$\fi}

\newcommand{  \mmdotlarge}{\ifmmode \dot{M}_*/\Mbhdot \else $\dot{M}_*/\Mbhdot$\fi}
\newcommand{  \mmdotsmall}{\ifmmode \Mbhdot/\dot{M}_* \else $\Mbhdot/\dot{M}_*$\fi}

\newcommand{  \mmwp     }{\ifmmode \left(M_{*}/M_{\rm BH}\right) \else $\left(M_{*}/M_{\rm BH}\right)$\fi}
\newcommand{  \ml       }{\ifmmode M_{*}/L_{*} \else $M_{*}/L_{*}$\fi}
\newcommand{  \mlwp     }{\ifmmode \left(M_{*}/L\right) \else $\left(M_{*}/L\right)$\fi}
\newcommand{  \mlk      }{\ifmmode \left(M_{*}/L_{K}\right) \else $\left(M_{*}/L_{K}\right)$\fi}
\newcommand{  \sigs     }{\ifmmode \sigma_{*} \else $\sigma_{*}$\fi}
\newcommand{  \Reff     }{\ifmmode R_{\rm e} \else $R_{\rm e}$\fi}
\newcommand{  \Rvir     }{\ifmmode R_{\rm vir} \else $R_{\rm vir}$\fi}
\newcommand{  \Rtwo     }{\ifmmode R_{200} \else $R_{200}$\fi}
\newcommand{  \Rfive    }{\ifmmode R_{500} \else $R_{500}$\fi}
\newcommand{  \Rgrp     }{\ifmmode R_{\rm grp} \else $R_{\rm grp}$\fi}
\newcommand{  \nser     }{\ifmmode n_{\rm s} \else $n_{\rm s}$\fi}
\newcommand{  \LSF      }{\ifmmode L_{\rm SF}  \else $L_{\rm SF}$\fi}
\newcommand{  \LFIR     }{\ifmmode L_{\rm FIR} \else $L_{\rm FIR}$\fi}
\newcommand{  \Lfir     }{\ifmmode L_{\rm FIR} \else $L_{\rm FIR}$\fi}
\newcommand{  \LTIR     }{\ifmmode L_{\rm TIR} \else $L_{\rm TIR}$\fi}
\newcommand{  \Ltir     }{\ifmmode L_{\rm TIR} \else $L_{\rm TIR}$\fi}

\newcommand{  \mdyn     }{\ifmmode M_{\rm dyn} \else $M_{\rm dyn}$\fi} 
\newcommand{  \mgas     }{\ifmmode M_{\rm gas} \else $M_{\rm gas}$\fi} 
\newcommand{  \mh       }{\ifmmode M_{\rm h} \else $M_{\rm h}$\fi}
\newcommand{  \mhalo    }{\ifmmode M_{\rm halo} \else $M_{\rm halo}$\fi}
\newcommand{  \sfr      }{\ifmmode {\rm SFR} \else SFR\fi}

\newcommand{ \Lcii     }{\ifmmode L_{\cii} \else $L_{\cii}$\fi}
\newcommand{ \fwcii  }{\ifmmode {\rm FWHM}\cii \else FWHM\cii\fi}

\newcommand{  \chandra }  {{\it Chandra}}
\newcommand{  \xmm     }  {{\it XMM-Newton}}

\newcommand{  \swift     }  {{\it Swift}}

\newcommand{\bj}{\ifmmode b_{\rm J} \else $b_{\rm J}$\fi}

\newcommand{\iab}{\ifmmode i_{\rm AB} \else $i_{\rm AB}$\fi}

\newcommand{\jab}{\ifmmode J_{\rm AB} \else $J_{\rm AB}$\fi}
\newcommand{\hab}{\ifmmode H_{\rm AB} \else $H_{\rm AB}$\fi}
\newcommand{\kab}{\ifmmode K_{\rm AB} \else $K_{\rm AB}$\fi}

\newcommand{\jveg}{\ifmmode J_{\rm Vega} \else $J_{\rm Vega}$\fi}
\newcommand{\hveg}{\ifmmode H_{\rm Vega} \else $H_{\rm Vega}$\fi}
\newcommand{\kveg}{\ifmmode K_{\rm Vega} \else $K_{\rm Vega}$\fi}

\newcommand{  \Chisq    }{\ifmmode \chi^{2} \else $\chi^{2}$}
\newcommand{  \nelec    }{\ifmmode n_{e} \else $n_{e}$\fi}     
\newcommand{  \nh       }{\ifmmode n_{\rm H} \else $n_{\rm H}$\fi}     
\newcommand{  \Ncol     }{\ifmmode N_{\rm col} \else $N_{\rm col}$\fi} 
\newcommand{  \NH       }{\ifmmode N_{\rm H} \else $N_{\rm H}$\fi}     

\def\sun{\hbox{$\odot$}}
\def\ion#1#2{#1$\;${\small\rm\@Roman{#2}}\relax}

\newcommand{\Lbat}{\ifmmode L_{\rm BAT} \else $L_{\rm BAT}$\fi}

\newcommand{\gamx}{\ifmmode \Gamma_{\rm x} \else $\Gamma_{\rm x}$\fi}
\newcommand{\gamtot}{\ifmmode \Gamma_{\rm tot} \else $\Gamma_{\rm tot}$\fi}
\newcommand{\gamsoft}{\ifmmode \Gamma_{\rm 0.3-10} \else $\Gamma_{\rm 0.3-10}$\fi}
\newcommand{\gamnec}{\ifmmode \Gamma_{\rm nEc} \else $\Gamma_{\rm nEc}$\fi}
\newcommand{\gambat}{\ifmmode \Gamma_{\rm BAT} \else $\Gamma_{\rm BAT}$\fi}
\newcommand{\gamref}{\ifmmode \Gamma_{\rm ref} \else $\Gamma_{\rm ref}$\fi}
\newcommand{\gamsim}{\ifmmode \Gamma_{\rm simple} \else $\Gamma_{\rm simple}$\fi}
\newcommand{\Ncnts}{\ifmmode N_{\rm counts} \else $N_{\rm counts}$\fi}

\newcommand{\Nswift}{1210}
\newcommand{\NswAGN}{836} 
 
\newcommand{\Nbassdrone}{642}
\newcommand{\Nzbass}{580}
\newcommand{\Nzned}{62}
\newcommand{\Nraw}{425}
\newcommand{\Nstart}{425}
\newcommand{\Ngood}{228}
\newcommand{\Nmbhdirect}{30}
\newcommand{\Nmbhhb}{126}
\newcommand{\Nmbhha}{23}
\newcommand{\Nmbhse}{149}
\newcommand{\Nmbhsigs}{49}
\newcommand{\Ntypeone}{174}
\newcommand{\Ntypetwo}{54}
\newcommand{\Nunobsc}{162}
\newcommand{\Nobsc}{66}
\newcommand{\Nobschi}{27}
\newcommand{\Ncthick}{8}

\title[BASS VI: the $\gamx-\lledd$ relation]{The {\it Swift}/BAT AGN Spectroscopic Survey (BASS) -- VI.\\ The $\Gamma_{\rm X} - L/L_{\rm Edd}$ relation}

\author[B. Trakhtenbrot et al.]{Benny Trakhtenbrot,$^{1}$\thanks{Zwicky fellow. E-mail: benny.trakhtenbrot@phys.ethz.ch}
Claudio Ricci,$^{2,3}$
Michael J. Koss,$^{1,4}$\thanks{Ambizione Fellow.}
Kevin Schawinski,$^{1}$
\newauthor
Richard Mushotzky,$^{5}$
Yoshihiro Ueda,$^{6}$
Sylvain Veilleux,$^{5}$
Isabella Lamperti,$^{1,7}$
\newauthor
Kyuseok Oh,$^{1}$
Ezequiel Treister,$^{2}$
Daniel Stern,$^{8}$
Fiona Harrison,$^{9}$
\newauthor
Mislav Balokovic,$^{9}$ 
and Neil Gehrels$^{10}$\thanks{Deceased 2017 February 6. This work is dedicated to his memory.}\\
$^{1}$ Institute for Astronomy, Department of Physics, ETH Zurich,Wolfgang-Pauli-Strasse 27, CH-8093 Zurich, Switzerland\\
$^{2}$ Instituto de Astrof\'{\i}sica, Facultad de F\'{\i}sica, Pontificia Universidad Cat\'olica de Chile, Casilla 306, Santiago 22, Chile\\
$^{3}$ Kavli Institute for Astronomy and Astrophysics, Peking University, Beijing 100871, China\\
$^{4}$ Eureka Scientific Inc., 2452 Delmer St. Suite 100, Oakland, CA 94602, USA\\
$^{5}$ Department of Astronomy and Joint Space-Science Institute, University of Maryland, College Park, MD 20742, USA\\
$^{6}$ Department of Astronomy, Kyoto University, Kyoto 606-8502, Japan\\
$^{7}$ Astrophysics Group, Department of Physics and Astronomy, University College London, 132 Hampstead Road, London NW1 2PS, UK\\
$^{8}$ Jet Propulsion Laboratory, California Institute of Technology, 4800 Oak Grove Drive, MS 169-224, Pasadena, CA 91109, USA\\
$^{9}$ Cahill Center for Astronomy and Astrophysics, California Institute of Technology, Pasadena, CA 91125, USA\\
$^{10}$ NASA Goddard Space Flight Center, Greenbelt, MD 20771, USA
}

\date{Accepted XXX. Received YYY; in original form \today}

\pubyear{2017}

\begin{document}
\label{firstpage}
\pagerange{\pageref{firstpage}--\pageref{lastpage}}
\maketitle

\begin{abstract}
We study the relation between accretion rate (in terms of \lledd) and shape of the hard X-ray spectral energy distribution (namely the photon index \gamx) for a large sample of 228 hard X-ray selected, low-redshift active galactic nuclei (AGN), drawn from the \swift/BAT AGN Spectroscopic Survey (BASS).
This includes 30 AGN for which black hole mass (and therefore \lledd) is measured directly through masers, spatially resolved gas or stellar dynamics, or reverberation mapping. 
The high quality and broad energy coverage of the data provided through BASS allow us to examine several alternative determinations of both \gamx\ and \lledd.
For the BASS sample as a whole, we find a statistically significant, albeit very weak correlation between \gamx\ and \lledd.
The best-fitting relations we find, $\gamx\simeq0.15\log\lledd+{\rm const.}$, are considerably shallower than those reported in previous studies.
Moreover, we find \emph{no} corresponding correlations among the subsets of AGN with different \mbh\ determination methodology. 
In particular, we find no robust evidence for a correlation when considering only those AGN with direct or single-epoch \mbh\ estimates. 
This latter finding is in contrast to several previous studies which focused on $z>0.5$ broad-line AGN. 
We discuss this tension and conclude that it can be partially accounted for if one adopts a simplified,  power-law X-ray spectral model, combined with \lledd\ estimates that are based on the continuum emission and on single-epoch broad line spectroscopy in the \emph{optical} regime.
We finally highlight the limitations on using \gamx\ as a probe of supermassive black hole evolution in deep extragalactic X-ray surveys.
%
\end{abstract}


\clearpage
\newpage

\section{Introduction}
\label{sec:intro}

One of the major goals in the study of active galactic nuclei (AGN) is to understand how basic physical properties of the accreting supermassive black hole (SMBH) are linked to the emergent (continuum) radiation field.
The UV-optical continuum can be explained by (thin) accretion discs, in a way which involves the BH mass (\mbh), accretion rate (in terms of the Eddington ratio, \lledd), and the BH spin, through deterministic, analytical and/or numerical models \cite[e.g.,][and references therein]{DavisHubney2006,Done2012,Netzer2013_book}.
This is not the case with the X-ray continuum emission, which is thought to originate from a compact, hot corona that surrounds the inner parts of the accretion disc and Compton up-scatters the disc UV photons. 
Indeed, it is not yet clear whether this significant emission component can be directly linked to any key AGN properties, from both theoretical and observational perspectives.

In the energy range $\sim0.5-10\,\kev$ the intrinsic X-ray continuum emission is observed to follow a power-law of the form $dN/dE\propto E^{-\gamx}$. 
Early evidence for a correlation between \gamx\ and \lledd\ was put forward by several studies that focused mainly on specific high-\gamx\ AGN and/or on narrow-line Seyfert 1 sources \cite[e.g.,][]{Pounds1995_RE1034,Brandt1997_Gamma_NLSy1s,BrandtBoller1998_NLSy1s,Porquet2004_XMM_PGs,Wang2004_Xcorona,Bian2005_GammaX_LLedd}. 
These sources, which are generally thought to represent the high-\lledd\ end of the (local) AGN population, exhibit soft X-ray spectra (i.e., high \gamx).
However, the limited size and range of luminosities probed in these early studies prohibited them from ruling out a scenario where the fundamental underlying relation is driven by \Lagn\ (or \mbh), rather than \lledd.

Since then, several studies have provided an increasingly more complete picture of this proposed relation by probing AGN that cover a wide range of luminosities and redshifts. 
These include the studies of 
\cite{Shemmer2006_Gamma_LLedd} and \citet[][S08 hereafter]{Shemmer2008_Gamma_LLedd}, which were the first to provide measurements for a substantial sample of $z>3$, extremely luminous quasars; 
\citet[R09 hereafter]{Risaliti2009_Gamma_LLedd}, which relied on a large sample of quasars drawn from the Sloan Digital Sky Survey (SDSS), at $0\ltsim z \ltsim 4.5$; 
\citet[B13 hereafter]{Brightman2013_Gamma_Ledd}, which used dozens of un-obscured, moderate-luminosity AGN at $0.5\ltsim z \ltsim 2$, from the COSMOS survey;
and \cite{Fanali2013_Gamma_LLedd}, which studied a  sample of un-obscured AGN from the \xmm\ Bright Serendipitous Survey.
The common result of these studies is the identification of a robust, statistically significant positive correlation between \gamx, usually measured over the observed-frame range of $\sim2-10\,\kev$, and \lledd.
Moreover, these studies demonstrated that \lledd\ is indeed the main driver of this relation, and not \Lagn\ and/or \mbh.
Most recently, the study of \cite{Brightman2016_Gamma_masers} showed that the $\gamx-\lledd$ relation is also applicable to local Compton-thick (CT) AGN, using a small sample of sources for which precise, maser-based determinations of \mbh\ are available.
Finally, several X-ray variability studies identified a trend of increasing \gamx\ with increasing flux levels for individual systems \cite[e.g,][]{Magdziarz1998_NGC5548,Zdziarski2003_Xray_radio,Sobolewska2009_Xray_var}, which may be interpreted as the consequence of a strong, positive correlation between \lledd\ and \gamx\ for any given accreting BH.

Besides the implications for small-scale physics of the X-ray emitting coronae near accreting SMBHs, S08 and the studies that followed highlighted the prospect of using \gamx, measured in deep X-ray surveys extending to $z\sim5$ \cite[e.g.,][]{BrandtAlexander2015_Rev}, to study the distribution of \lledd\ and hence of \mbh\ (assuming a bolometric correction), in virtually \emph{all} classes of AGN, including obscured sources where \lledd\ and \mbh\ are otherwise unavailable.
Moreover, it has been shown that a positive relation between \gamx and \lledd\ would be able to explain the X-ray Baldwin effect \cite[][]{Ricci2013_LLedd_Gamma_Ka}, i.e. the decrease of the Fe\,K$\alpha$ equivalent width with increasing luminosity and \lledd\ \cite[e.g.,][]{IwasawaTaniguchi1993_Ka_BE,Bianchi2007_Ka_BE}.

A possible explanation of the $\gamx-\lledd$ relation is that for high \lledd\ the intense UV/optical radiation, which provides the seed photons for the X-ray emission, can lead to a more efficient cooling of the X-ray corona, decreasing the temperature and/or optical depth of the plasma. 
Lower temperatures and/or optical depths of the corona, in turn, would then result in a softer X-ray SED (i.e., higher \gamx). 
This would also be in agreement with the positive correlation between \lledd\ and the ratio of the UV/optical-to-X-ray flux found in several studies 
(see, e.g., S08; \citealt{Grupe2010_SED,Lusso2010_XCOS_type1,Jin2012_Xop_3_SEDs}, 
but also \citealt{VasudevanFabian2007_BC,Vasudevan2009_Xray_opt_BAT9m}).
Another explanation involves the growth of instabilities in a two-phase, disc-corona accretion flow, which would also explain the flattening (or, indeed, reversal; see below) of the $\gamx-\lledd$ relation at low accretion rates \cite[][]{Yang2015_Gamma_LLedd,Kawamuro2016_lowL_BAT_Suzaku}.

It is worth bearing in mind, however, that virtually all the studies that reported $\gamx-\lledd$ correlations used a relatively limited X-ray energy range, dictated by the capabilities of the \chandra\ and \xmm\ facilities (i.e., $\sim0.5-10\,\kev$).
One may suspect that this limited energy range may not be broad enough to account for the rich collection of phenomenological and physical radiation components which constitute the X-ray SED of AGN.
In particular, some studies noted the issues involving the soft excess, the Compton ``hump'', and/or the high energy cut-off, and how the inability to observe these directly may affect the spectral decomposition of the AGN samples under study.
Other limitations of the reported positive $\gamx-\lledd$ correlation include the significant amount of scatter observed in the $\gamx-\lledd$ plane, where \gamx\ may cover the range $\gamx\sim1-2.5$ even for a narrow range in \lledd\ \cite[$\lesssim 0.5$ dex; see, e.g.,][]{HoKim2016_lowM_X_var}.
Finally, some studies have claimed that, for slowly accreting and/or low mass SMBHs, the $\gamx-\lledd$ relation may actually change to become an \emph{anti-}correlation \cite[e.g.,][]{Constantin2009_LLAGN_ChaMP,Younes2011_LINERS,Kamizasa2012_IMBHs_X_var,Yang2015_Gamma_LLedd,Kawamuro2016_lowL_BAT_Suzaku}.

A promising way for addressing some of these limitations, and for expanding the $\gamx-\lledd$ relation towards more complete, larger samples of AGN, is to study hard X-ray selected AGN, for which the spectral coverage in the X-rays extends to higher energies.
Indeed, several studies tried to identify relations between \gamx\ and \lledd\ in samples of AGN detected by the hard X-ray \swift/BAT instrument (covering roughly $15-150\,\kev$; \citealt{Gehrels2004_Swift}), which are  essentially free of any obscuration-related selection biases.
Some of these earlier \swift/BAT studies demonstrated the significant scatter in the $\gamx-\lledd$ plane, and found no convincing evidence for a correlation between these properties -- interpreted as a result of the limited sample size \cite[e.g.,][]{Winter2009_BAT_Xray_SEDs,Winter2009_BAT_Suzaku_var}.
The increasing size of \swift/BAT-detected AGN samples, combined with more elaborate X-ray spectral analyses, eventually allowed the identification of significant $\gamx-\lledd$ correlations (\citealt{Winter2012_X_SEDs_WAs,Kawamuro2016_Suzaku}; note that in the former study the correlations are found only when binning the sample by \lledd, similarly to B13).

In the present study, we seek to establish a relation between \gamx\ and \lledd\ for a large and essentially complete sample of low-redshift, hard X-ray selected AGN.
Our sample is based on the first data release of the \swift/BAT AGN Spectroscopic Survey (BASS).
BASS provides a rich collection of X-ray and optical data for about \Nbassdrone\ AGN, mostly at $z<0.5$, with unprecedented levels of completeness in terms of optical spectroscopy. 
Compared to other low-redshift AGN samples, the hard X-ray selection that forms the basis of BASS ensures that the resulting sample is minimally affected by the AGN hosts, particularly by obscuring dust and/or contaminating optical line emission.
The BASS sample covers a wide range in \Lbol, \mbh, \lledd, for AGN of essentially all emission line and/or obscuration based classification.
It therefore serves as an ideal benchmark for addressing many open questions concerning the X-ray and optical emission mechanisms in AGN, and how these are related to basic BH properties \cite[][]{Berney2015_BASS,Lamperti2017_BASS_NIR,Oh2017_BASS_LLedd_BPT}.

\begin{figure*}
\includegraphics[width=0.32\textwidth]{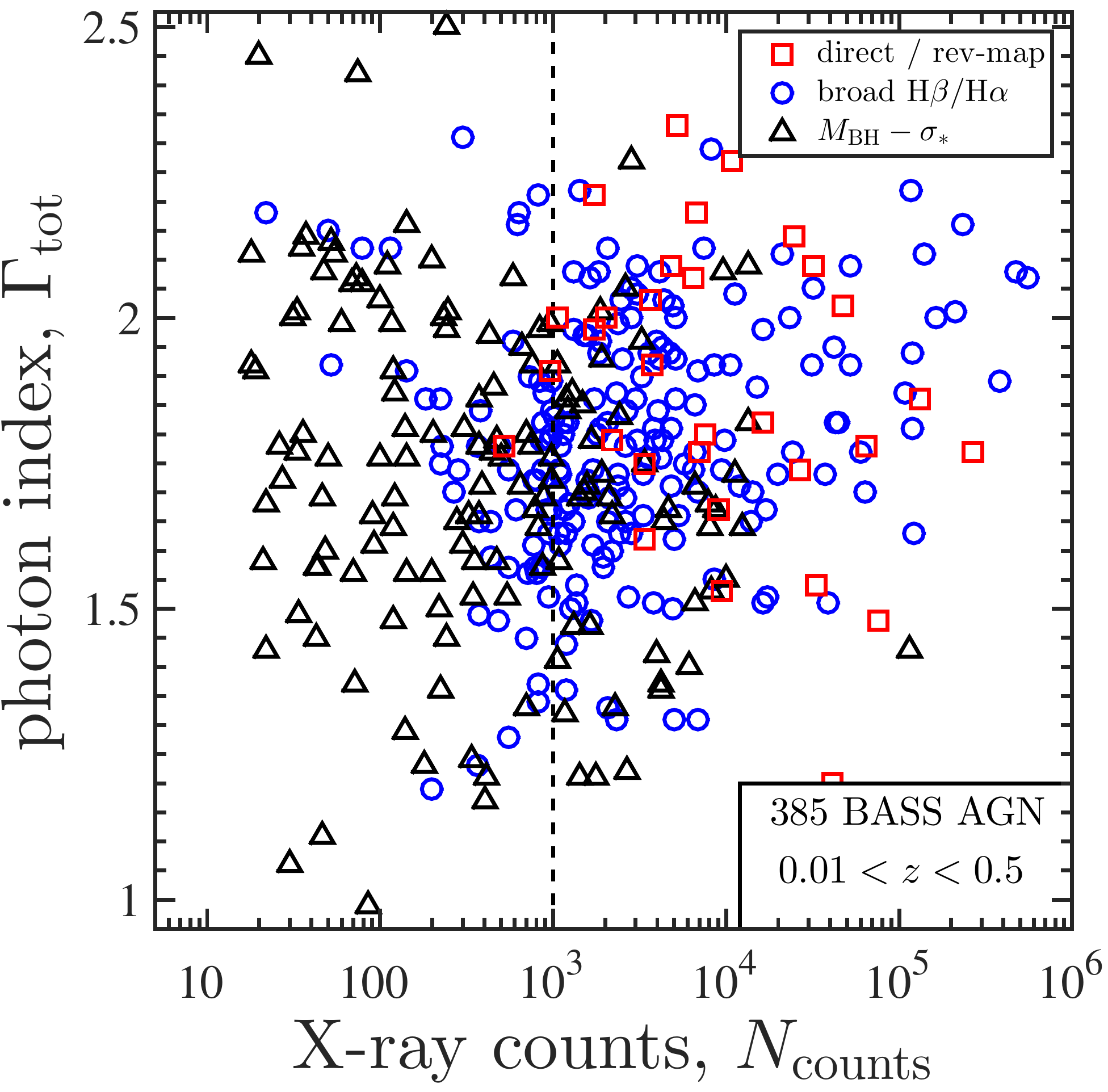}
\hfill
\includegraphics[width=0.32\textwidth]{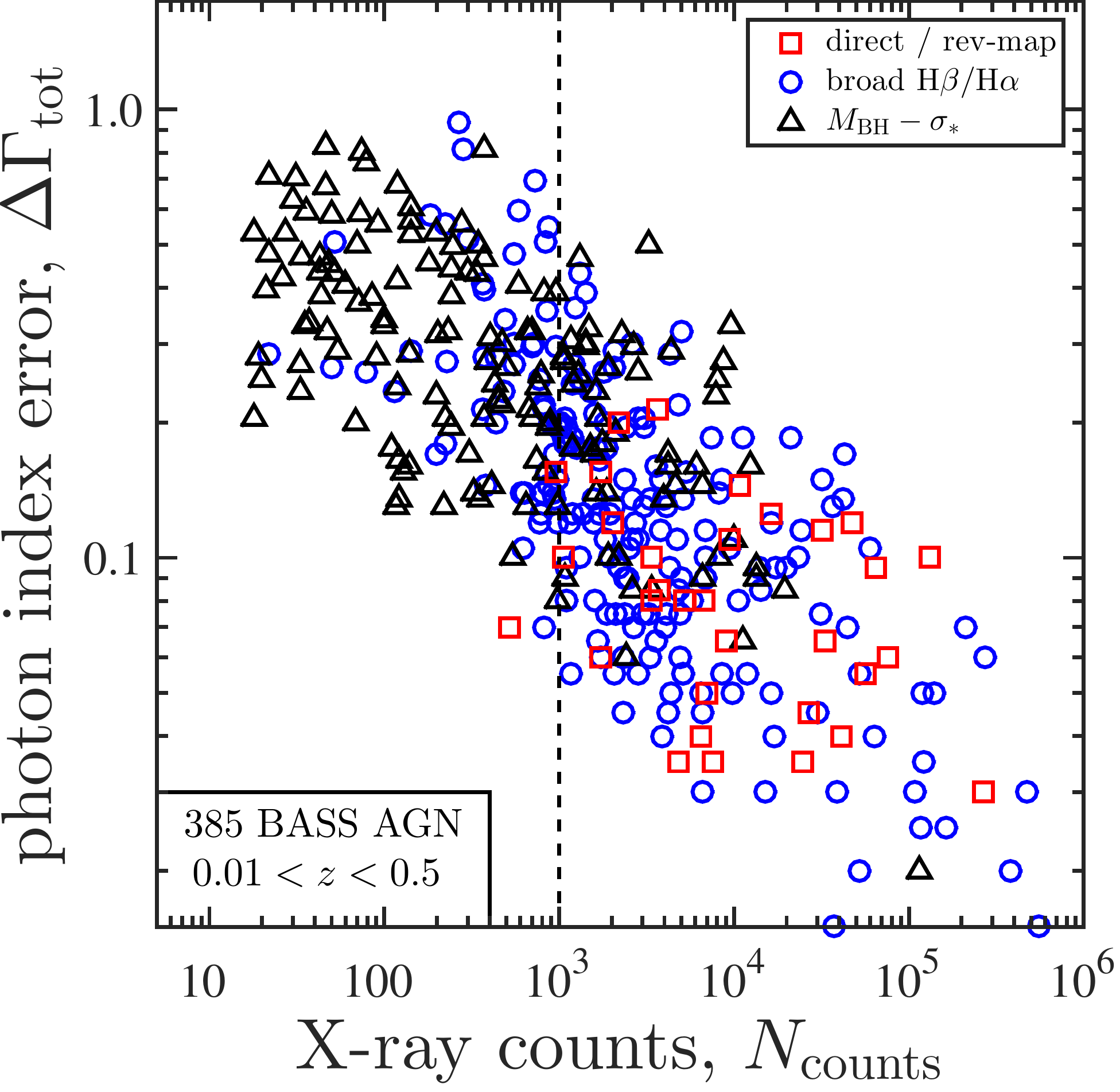}
\hfill
\includegraphics[width=0.32\textwidth]{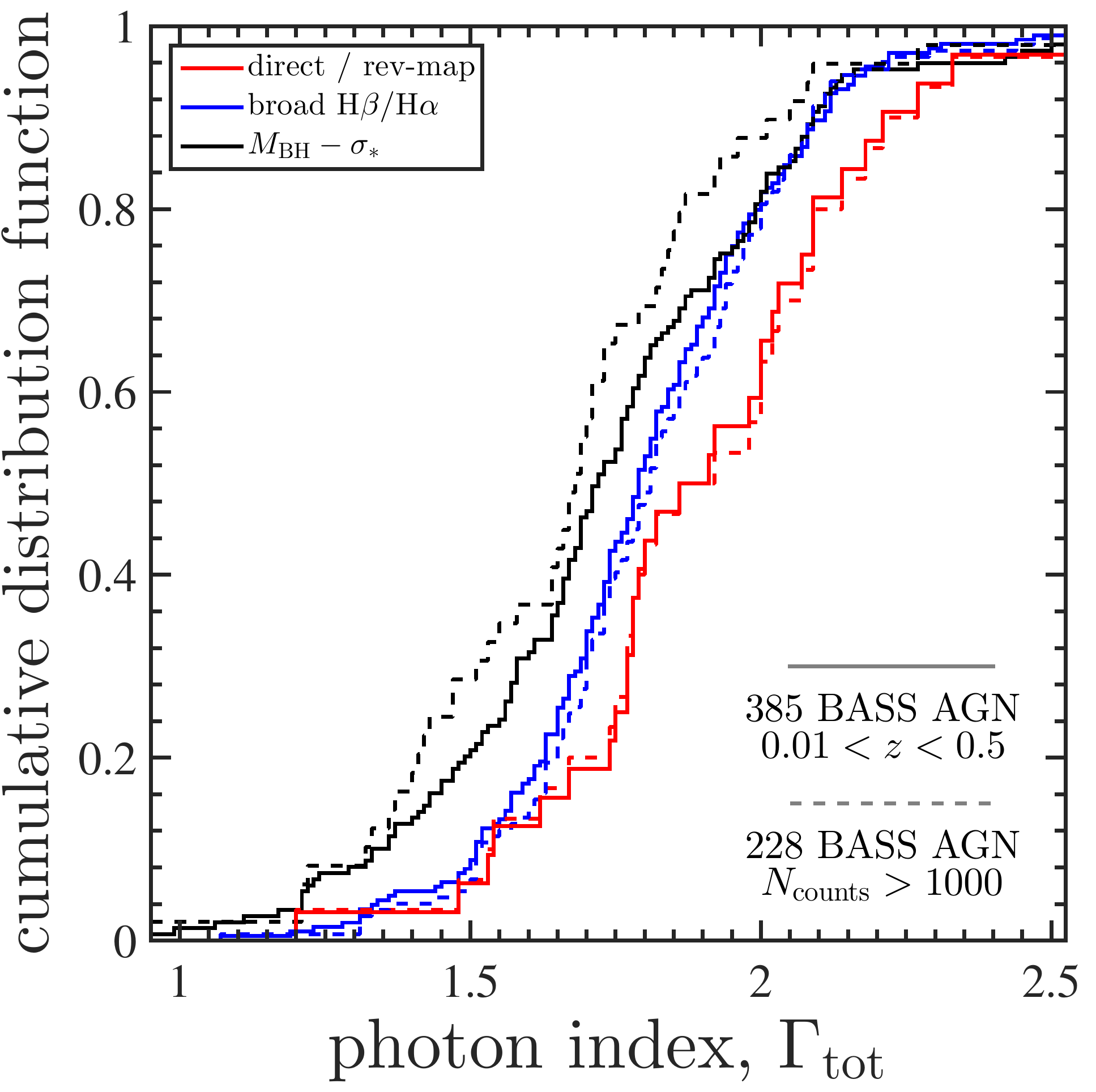}
\caption{
Statistical properties of the hard X-ray photon index, \gamx, in our AGN sample.
{\it Left}: The best-fitting photon index over the entire available range of X-ray data, \gamtot, vs.\ the number of counts in this energy range, \Ncnts. 
Our AGN are split according to the \mbh\ (and therefore \lledd) determination method: 
either through ``direct'' methods (masers, resolved gas or stellar kinematics, or reverberation; red squares); through ``single-epoch'' mass estimators using one of the broad Balmer lines (blue circles); or through stellar velocity dispersions (\sigs) and the $\mbh-\sigs$ relation (black triangles).
The dashed vertical line denotes $\Ncnts=1000$ - the conservative threshold we apply to the parent BASS sample to focus on high-quality X-ray data.  
{\it Centre}: the uncertainty on the photon index, $\Delta\gamtot$, vs.\ \Ncnts.
{\it Right}: cumulative distribution function of \gamtot. Lines of different colours trace the \mbh\ subsets.
For each subset, solid and dashed lines trace the distribution of \gamtot\ among all AGN and among those with $\Ncnts>1000$, respectively.
}
\label{fig:gamma_stats}
\end{figure*}

This paper is organized as follows.
In Section~\ref{sec:sample_data} we present our sample and the data from which we measure \lledd\ and \gamx.
In Section~\ref{sec:results} we examine possible correlations between these quantities, but conclude that robust and/or strong correlations of this sort cannot be clearly established for our BASS sample of AGN.
In Section~\ref{sec:discuss} we discuss our main findings, in the context of the several previous studies that reported $\gamx-\lledd$ relations.
Section~\ref{sec:conclusions} summarises our findings.
Throughout this work we assume a cosmological model with $\Omega_{\Lambda}=0.7$, $\Omega_{\rm M}=0.3$, and $H_{0}=70\,\kms\,{\rm Mpc}^{-1}$.

\section{Sample and Data}
\label{sec:sample_data}

\subsection{The BASS DR1 Sample of AGN}
\label{subsec:sample}

This work focuses on AGN selected through their hard-band X-ray emission, as identified in the \swift/BAT 70-month catalogue \cite[][]{Baumgartner2013_SwiftBAT_70m}.
Out of \Nswift\ unique objects in that catalogue, \NswAGN\ have been identified as known AGN.
The first data release of the BASS project (Koss et al., submitted; K17 hereafter) includes \Nbassdrone\ of these AGN, for which redshifts and complementary multi-wavelength data is available.
As part of the BASS effort, we have curated optical spectra for \Nzbass\ AGN, which were then used to measure accurate redshifts and other spectral properties, relying on narrow emission lines (usually \OIII; see K17). 
For \Nzned\ additional AGN, redshifts are available from the NASA Extragalactic Database (NED).

Of the initial sample of \Nbassdrone\ sources with redshifts, we first focus on the \Nstart\ AGN for which determinations of \mbh\ are available within BASS.
More information regarding these determinations of \mbh\ is provided in Section~\ref{subsec:Lbol_lledd} below.
We further focus on sources within the redshift range $0.01<z<0.5$, thus omitting 40 AGN and leaving in 385 BASS sources.
This is done to avoid high-$z$ beamed AGN (and blazars), and also extremely nearby AGN for which the precise distances (and therefore luminosities) may be somewhat uncertain.
We finally select only those sources for which the available X-ray observations have a sufficiently high number of counts, i.e. $\Ncnts>1000$, to ensure a high quality spectral fit (more information regarding our X-ray data is given in Section~\ref{subsec:xray_data} below).

Our final primary sample therefore consists of \Ngood\ AGN at $0.01<z<0.5$, which have a high-quality X-ray spectrum and a reliable BH mass determination. 
These include 
\Nmbhdirect\ AGN with ``direct'' mass measurements (either from masers, gas- or stellar-dynamics, or reverberation mapping); 
\Nmbhse\ AGN with \mbh\ estimates obtained through single-epoch spectra of broad Balmer lines; and 
\Nmbhsigs\ AGN for which \mbh\ is determined by combining stellar velocity dispersion (\sigs) measurements and the $\mbh-\sigs$ relation. 
We note that -- unlike other samples that investigated the relation between \gamx\ and \lledd\ -- our final sample consists of both broad- and narrow-line AGN (\Ntypeone\ and \Ntypetwo\ AGN, respectively).

\subsection{X-ray Data and Analysis}
\label{subsec:xray_data}

The analysis of the available X-ray data for the BASS AGN was presented in detail in Ricci \et (submitted; R17 hereafter). 
This analysis included all the X-ray data available for the BASS sample, including \swift/XRT, \xmm/EPIC, \chandra/ACIS, {\it Suzaku}/XIS, or {\it ASCA}/GIS/SIS observations, and typically covering the observed-frame energy range of $0.3-150\,\kev$.
The X-ray data were fitted with a set of models that rely on an absorbed power-law X-ray SED with a high-energy cut-off, and a reflection component. 
A cross-calibration constant was applied to each source, in order to account for possible flux variability between the 70-month integrated \swift/BAT spectrum and the significantly shorter 0.3--10\,keV observations. 
Additional components accounting for warm absorbers, soft excess, Fe\,K$\alpha$ lines, and/or other spectral features were added if deemed necessary to obtain a satisfactory fit to the data. 
The reader is referred to R17 for a detailed discussion of the models' physical components, parameters, and fitting quality.
We note here that the R17 analysis did \emph{not} explicitly impose a finite range of possible \gamx. 

The analysis of the X-ray data provided several ways of determining \gamx, which we use throughout the present study:
\begin{itemize}
\item 
First, \gamtot\ denotes the photon index recovered from the entire (relevant) energy range and the full multi-component model adopted for each source (in which $E_{\rm C}$ is a free parameter).
This is the fiducial photon index adopted in the R17 study and throughout the present work (unless otherwise noted).

\item
\gamnec\ results from modelling the entire X-ray spectral range with a model which ignores the high-energy cut-off (i.e., setting $E_{\rm C}=500\,\kev$).

\item
\gamsoft\ denotes the photon index that describes only the observed-frame 0.3-10\,\kev\ energy range, using a model that ignores the high-energy cut-off (which was fixed to $E_{\rm C}=500\,\kev$) and the reflection component. 

\item
\gambat\ results from fitting a power-law model solely to the energy range probed by \swift/BAT (i.e., 14-195\,\kev).

\end{itemize}

Figure~\ref{fig:gamma_stats} presents some of the statistical properties of \gamtot\ for our sources.  
The left and centre panels show \gamtot\ and the related uncertainty ($\Delta\gamtot$), plotted against the number of counts across the available X-ray spectral range, \Ncnts. 
As noted above, in this work we include only BASS AGN with $\Ncnts>1000$, where the typical uncertainty on \gamtot\ is $\Delta\gamtot\lesssim0.2$.
This choice, which we can make only thanks to the high-quality X-ray data in BASS, can be considered conservative -- indeed, previous studies of the $\gamx-\lledd$ relation relied on X-ray spectra with significantly fewer counts.\footnote{
For example, only about $\sim1/3$ of those studied by B13 had $\Ncnts\gtrsim1000$ (a cut of $\Ncnts>250$ was applied for the entire B13 sample).} 
The left panel of Fig.~\ref{fig:gamma_stats} suggests that our cut on \Ncnts\ does not bias our sample against any particular range in \gamtot.

Our primary sample of \Ngood\ AGN with $\Ncnts>1000$ and $0.01<z<0.5$ covers a wide range in \gamtot\, with $\gamtot\sim1-2.8$, a median (and mean) value of $\langle\gamtot\rangle=1.8$, and a standard deviation of $\sigma(\gamtot)=0.27$.
The R17 analysis of the X-ray data for our sources also provides a more detailed and complete method for quantifying the obscuration towards the BASS AGN, based on the (Hydrogen) column density \NH.
Setting the threshold at $\log(\NH/\cmii)=22$ splits our primary sample to \Nunobsc\ unobscured and \Nobsc\ obscured AGN (i.e., with $\log[\NH/\cmii]$ below and above $22$, respectively, and still obeying the redshift and \Ncnts\ cuts described above). 
\Nobschi\ of the AGN in our primary sample are heavily obscured, with $\log(\NH/\cmii)\ge 23.5$, and \Ncthick\ of these are Compton-thick ($\log(\NH/\cmii)\ge24$; see also \citealt{Ricci2015_local_CT}).
%

\subsection{Bolometric luminosities and \lledd}
\label{subsec:Lbol_lledd}

The BH masses available for all our \Ngood\ BASS sources were determined through several different methods.
First, for \Nmbhdirect\ sources, we relied on directly measured \mbh\ - either from masers; spatially-resolved gas or stellar dynamics; or from reverberation mapping.
For \Nmbhse\ AGN, our \mbh\ estimates rely on single-epoch spectra of the broad Balmer emission lines, and prescriptions that fundamentally rely on the results of reverberation mapping campaigns.
In particular, for \Nmbhhb\ sources, we used the broad \Hbeta\ emission line and the adjacent continuum luminosity ($\Lop \equiv \lamLlam$[5100\AA), relying on the same line fitting procedure and \mbh\ estimator as in \cite{TrakhtNetzer2012_Mg2}.
For \Nmbhha\ additional sources, \mbh\ is determined from the broad \halpha\ emission line, following the procedure described in \cite{Oh2015_hidden_BLAGN} and the prescription of \cite{Greene_Ho_Ha_2005}.
Finally, for \Nmbhsigs\ sources with no broad emission lines, we used \sigs\ measurements and the $\mbh-\sigs$ relation of \cite{KormendyHo2013_MM_Rev}.

As explained in the BASS/DR1 paper (K17), we prefer to use the ``direct'' \mbh\ determinations, whenever available. Otherwise, we use the single-epoch estimates from broad Balmer lines, and finally those from \sigs.
This reflects the different levels of uncertainty related to each of the mass estimation methods, which are discussed in K17.
We briefly note here that the uncertainties on BH masses derived through single-epoch spectra of broad lines -- which constitute the largest subset in our BASS sample -- may reach $\sim0.3-0.4$ dex \cite[see, e.g.,][and references therein]{Shen_Liu_2012,Shen2013_rev,Peterson2014_review,Mejia2016_XS_MBH}.
On the other hand, for \mbh\ determinations based on resolved stellar or gas dynamics (including masers), the statistical uncertainties are much lower, $\lesssim 0.1$ dex.
Importantly, the single-epoch mass estimators are calibrated in a way that minimizes  any systematic offsets with respect to other methods \cite[see, e.g.,][]{Park_LAMP_2012,Grier2013_sigs_PGs,Woo2013_RM_Msig}.

We estimated the bolometric luminosities of our sources, \Lbol, following several different prescriptions, based on the available X-ray and optical luminosities of our AGN.
We mainly use the (absorption-corrected) luminosities in the 2-10\,\kev\ rest-frame energy range, \Lhard, derived from the best-fitting, multi-component spectral models of the X-ray data (but ignoring any cross-calibration scaling factors; see R17).
These are combined with three different bolometric corrections.
First, we used a fixed bolometric correction of $\fbolhard\equiv\Lbol/\Lhard=20$, a typical value for AGN \cite[see, e.g.,][]{Elvis1994,Marconi2004,VasudevanFabian2007_BC,Jin2012_Xop_3_SEDs}.
Second, we used the \Lhard-dependent bolometric corrections of \cite{Marconi2004}.
For the sample considered in this study, these are in the range of $\fbolhard=11-140$, with a median value of $\fbolhard=26.7$, and 80\% of the sources having $\fbolhard\simeq18-48$. 
The resulting \Lbol\ are therefore slightly larger than those obtained through $\fbolhard=20$, by 0.1 dex (median value; the standard deviation is 0.18 dex), but otherwise there are no significant systematic differences between the two.
We have also examined the effects that an \lledd-based bolometric correction would have on our results. 
For this, we relied on the results of \cite{VasudevanFabian2007_BC}, which provide $\fbolhard=20$ for $\lledd\le0.04$, $\fbolhard=70$ for $\lledd\ge0.4$, and follow $\fbolhard\propto\lledd^{0.54}$ over the range $0.04<\lledd<0.4$. 
The more recent study of \cite{Jin2012_Xop_3_SEDs} suggests a similar dependence of \fbol.
\footnote{
It has been suggested that the $\fbolhard-\lledd$ relation is itself a reflection of an underlying $\gamx-\lledd$ positive correlation \cite[e.g.,][]{Fanali2013_Gamma_LLedd}.
}
We stress that these prescriptions for \Lbol\ may provide markedly different values for individual sources, and therefore potentially affect any analysis of the $\gamx-\lledd$ plane. 
We indeed consider them all in our analysis (see Section~\ref{subsec:res_other_gam}).

We additionally used the absorption-corrected BAT luminosities, which cover the range $14-150\,\kev$, combined with a fixed bolometric correction of $\fbolbat\equiv\Lbol/\Lbat=8.5$. 
This bolometric correction is derived from the $\fbolhard=20$ one, by assuming a constant $\gamx=1.8$ -- similar to the median value our sample (see Fig.~\ref{fig:gamma_stats}), which corresponds to $\Lbat/\Lhard=2.35$.
These \Lbat-based estimates of \Lbol\ are generally in very good agreement with the fiducial, \Lhard-based ones. The median difference is 0.04 dex (with \Lbat-based estimates of \Lbol\ being slightly lower), and the standard deviation is 0.21 dex.

Finally, for the subset of \Nmbhhb\ AGN for which \mbh\ was determined from single-epoch spectroscopy of the broad \hbeta\ line, we derived an additional set of \Lbol\ estimates using \Lop-dependent bolometric corrections, \fbolopt, which are calibrated against the \cite{Marconi2004} ones \cite[see also][]{TrakhtNetzer2012_Mg2}.
In the range of \Lop\ covered by our BASS sample, these can be approximated by $\log\fbolopt=-0.084\times\log(\Lop/10^{44}\,\ergs) + 0.9$.
This set of \lledd\ estimates is mainly used to allow a direct comparison with the analysis presented in some previous studies of the $\gamx-\lledd$ relation (e.g., S08; see Section \ref{subsec:comparison}).

Throughout this work we focus mainly on the \Lhard-based determinations of \Lbol\ (and therefore \lledd). 
This choice is mainly motivated by our attempt to minimize the effects of source variability on our measurements and analysis.
The lower-energy X-ray data, which dominate the determination of \gamx\ (and, obviously, of \Lhard), were obtained through integrations that are much shorter than the \swift/BAT ones (which, in turn, represent typical flux levels over a period of $\sim$70 months).
The Eddington ratios of our AGN, which provide a dimensionless, \mbh-normalized measure of the accretion rate onto the SMBHs, are then calculated following $\lledd=\Lbol/(1.3\times10^{38}\,\mbh/\Msol)$.

\section{The $\gamx-\lledd$ Relation for \swift/BAT AGN}
\label{sec:results}

\subsection{Straightforward analysis with $\gamtot-\lledd$}
\label{subsec:res_gamtot}

\begin{figure}
\includegraphics[width=0.4825\textwidth]{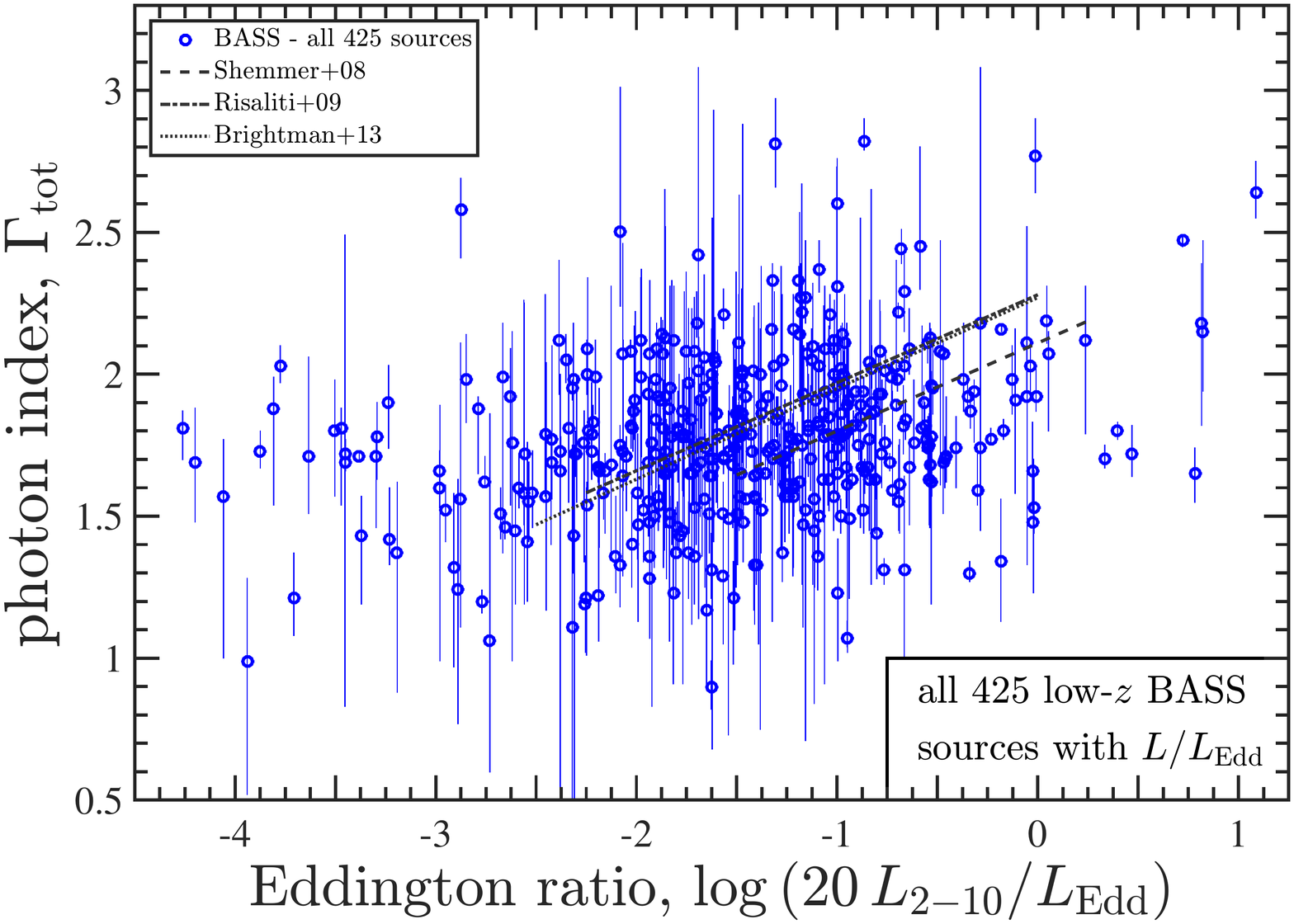}
\caption{
\gamtot\ vs.\ \lledd\ for the parent raw sample of \Nraw\ BASS AGN -- ignoring restrictions on the number of X-ray counts (\Ncnts) or on the lower redshift bound.
The grey diagonal lines represent the best-fitting $\gamx-\lledd$ relations reported by the studies of \citet{Shemmer2008_Gamma_LLedd}, \citet{Risaliti2009_Gamma_LLedd}, and \citet{Brightman2013_Gamma_Ledd}, over the (approximate) range covered by the respective samples.
}
\label{fig:gam_tot_lledd_raw}
\end{figure}

\begin{figure*}
\includegraphics[width=0.75\textwidth]{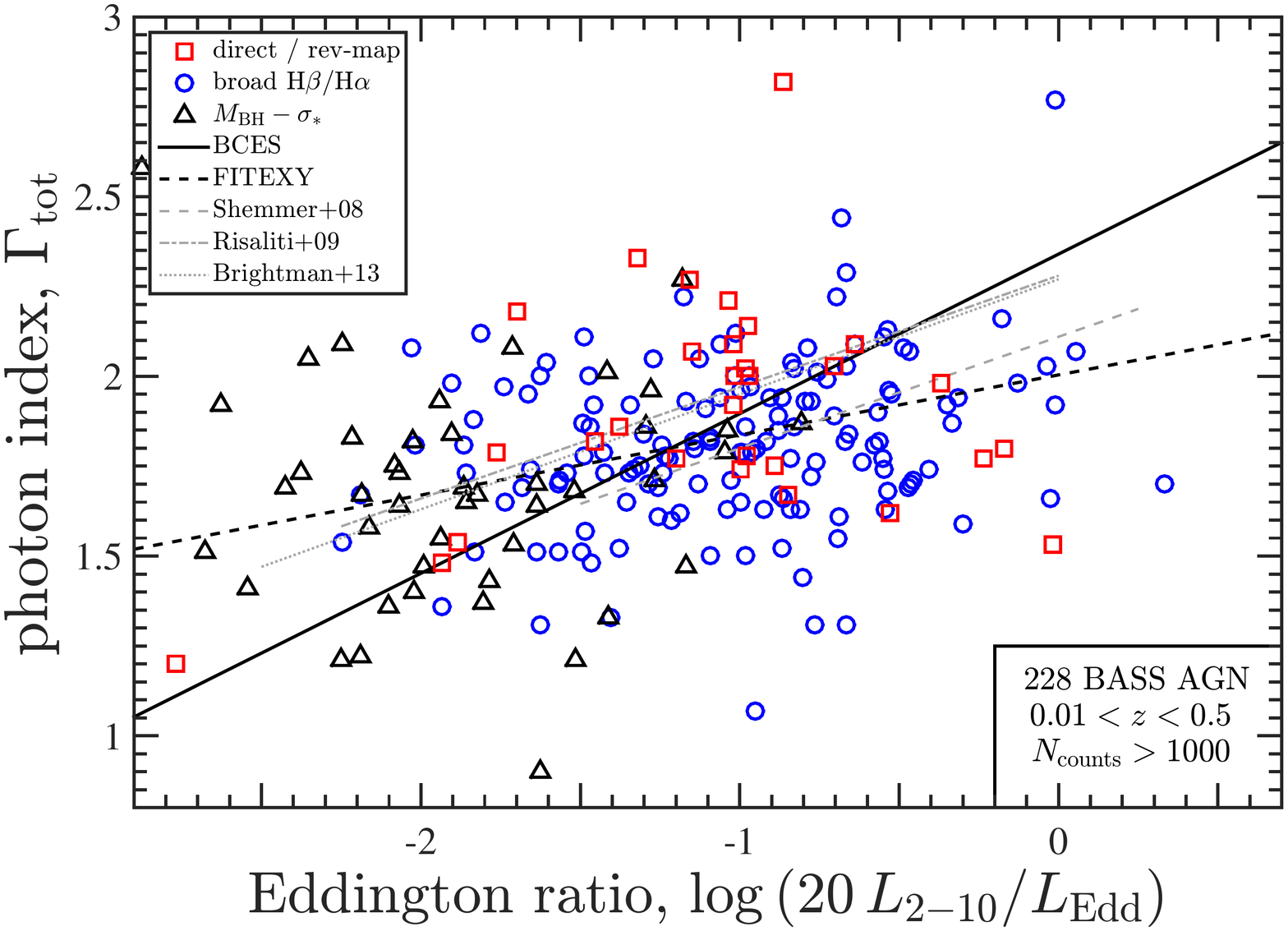}
\caption{
\gamtot\ vs.\ \lledd\ for our sample of \Ngood\ BASS AGN with $0.01<z<0.5$ and high-quality X-ray data ($\Ncnts>1000$).
Different symbols represent AGN for which \mbh\ (and therefore \lledd) is estimated either through ``direct'' methods (masers, resolved gas or stellar kinematics, or reverberation); through ``single-epoch'' mass estimators using one of the broad Balmer lines; or through stellar velocity dispersions (\sigs) and the $\mbh-\sigs$ relation.
The black diagonal lines represent the best-fitting $\gamtot-\lledd$ relations we obtain using either the \texttt{BCES} (bisector) or the \texttt{FITEXY} methods (solid and dashed lines, respectively).
The grey diagonal lines represent the best-fitting $\gamx-\lledd$ relations reported by the studies of \citet{Shemmer2008_Gamma_LLedd}, \citet{Risaliti2009_Gamma_LLedd}, and \citet{Brightman2013_Gamma_Ledd}, over the (approximate) range covered by the respective samples.
}
\label{fig:gam_tot_lledd_all}
\end{figure*}

Figure~\ref{fig:gam_tot_lledd_raw} shows the photon index vs. the accretion rate for the entire (parent) sample of \Nraw\ BASS AGN. 
We stress that this includes \emph{all} the non-blazar sources for which the quantities are available, ignoring (for now) the different redshift and \Ncnts\ cuts described above, and regardless of the method used for \mbh\ estimation.
Here we use the photon index we obtained from the entire spectral fit to the available X-ray data, \gamtot, and the \lledd\ estimates that are based on $\Lbol=20\times\Lhard$.

A formal (Spearman) hypothesis test results in a weak and only marginal statistically significant correlation between the quantities, with the probability of finding a correlation if the null hypothesis (i.e., no correlation) is true being  $P=0.8\%$, and a correlation coefficient of $r_{\rm s}=0.23$.\footnote{
Throughout this work we define a correlation as ``significant'' if the \emph{two-sided} Spearman correlation test results in $P<0.1\%$ (corresponding to $>3.3\sigma$). 
Correlations with $0.1<P<1\%$ (i.e., $\sim2.6-3.3\sigma$) are referred to as ``marginally significant'', in order to avoid a situation where small differences in $P$-values result in stark qualitative disagreements with previous works.
} 
Thus, it appears that our parent BASS sample may hold limited evidence for a $\gamx-\lledd$ relation of the kind found in several previous studies, although at lower statistical significance ($<3\sigma$).
However, in what follows we will demonstrate that this result is not robust, and in particular that it does not hold for subsets of sources that differ in the \mbh\ determination methodology, for alternative determinations of \lledd, and/or when some data quality cuts are imposed on the sample.


In Fig.~\ref{fig:gam_tot_lledd_all} we again show \gamtot\ vs. \lledd, but only for the \Ngood\ BASS AGN in our main sample, i.e. those that satisfy $\Ncnts>1000$ and $0.01<z<0.5$.
Here, too, we use $\Gamma_{\rm tot}$ and the \Lhard-based estimates of \lledd.
Fig.~\ref{fig:gam_tot_lledd_all} also shows the best-fit relations between \gamx\ and \lledd\ reported in the three main reference studies of S08, R09, and B13.
Adopting a notation of 
\begin{equation}
    \gamx = \alpha\,\log\left(\lledd\right) + \beta \,\, ,
	\label{eq:lledd_gamma}
\end{equation}
these studies have reported $\left(\alpha,\beta\right)=\left(0.31,2.11\right)$, $\left(0.31,2.28\right)$, and $\left(0.32,2.27\right)$, respectively.
\footnote{
For the R09 study, we list the relation which relies on the ``total'' sample, despite the fact that for $\sim$17\% of those AGN have \CIV-based determinations of \mbh\ (and therefore, \lledd), which are known to be problematic \cite[see][and references therein]{TrakhtNetzer2012_Mg2}.
The relation derived in R09 for AGN with \hb-based determinations of \mbh\ is much steeper, with $\left(\alpha,\beta\right)=\left(0.58,2.57\right)$.
}
The samples and methods used in these studies are described in Section~\ref{subsec:comparison}.

\begin{figure}
\includegraphics[width=0.475\textwidth]{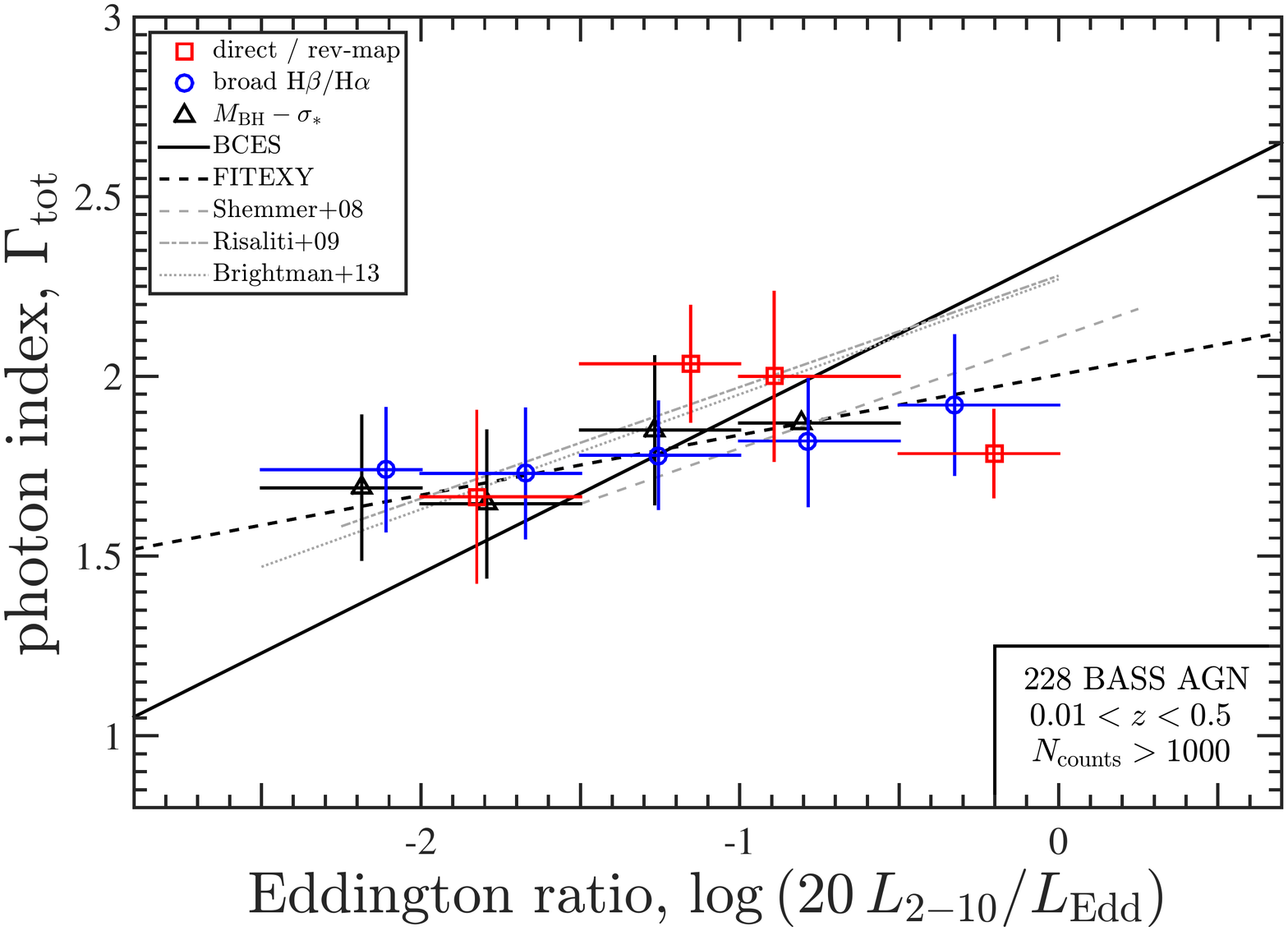}
\caption{
Same as Fig.~\ref{fig:gam_tot_lledd_all}, but binning the AGN in each \mbh\ subset according to \lledd, in steps of 0.5 dex.
Markers represent the median \lledd\ and \gamtot\ in each such bin. Error bars on \lledd\ represent the bin size, while those on \gamx\ represent the median absolute deviations (MAD).
All symbols and lines are identical to those in Fig.~\ref{fig:gam_tot_lledd_all} (the black lines represent the best-fitting relations obtained for the un-binned data).
}
\label{fig:gam_tot_lledd_binned}
\end{figure}

As Fig.~\ref{fig:gam_tot_lledd_all} clearly shows, there is a considerable amount of scatter and little evidence for strong trends between \gamtot\ and \lledd\ in our sample of \Ngood\ BASS AGN.
In an attempt to illustrate the overall trends that may be present in our sample, in Fig.~\ref{fig:gam_tot_lledd_binned} we show the \emph{binned} \gamx\ vs. \lledd\ for each of the \mbh\ subsets, where the bins spread 0.5 dex in \lledd. 
The markers represent the median values within each bin, while the vertical error bars represent the median absolute deviations (MAD) of \gamtot.
Fig.~\ref{fig:gam_tot_lledd_binned} further demonstrates the large scatter in the (underlying) BASS sample, and the limited evidence for a strong $\gamx-\lledd$ correlation for our AGN.
A formal correlation test does indeed show evidence for a weak, but statistically significant correlation: the null hypothesis of no correlation between \gamtot\ and \lledd\ can be rejected at a level corresponding to $P=1.65\times 10^{-4}\,\%$, when the entire sample of 228 AGN is considered. 
The corresponding Spearman correlation coefficient is $r_{\rm s}=0.31$ -- implying a weak correlation.
\footnote{We stress that this value of the correlation coefficient $r_{\rm s}$ should \emph{not} be directly compared with the slopes of the $\gamx-\lledd$ relations reported by the aforementioned studies, despite their similarity.}

We employ several linear regression analysis methods to derive the best-fitting parameters of the $\gamtot-\lledd$ correlation for the primary BASS sample.
In all these fits we assume a uniform uncertainty of 0.3 dex on \lledd\ (following S08).
The \texttt{BCES}(Y|X) method \cite[][]{Akritas1996_BCES} provides
\begin{equation}
    \gamtot = \left(0.167\pm0.04\right)\,\log\left(\lledd\right) + \left(2.00\pm0.05\right) \,\, ,
	\label{eq:lledd_gamtot_bces_bis}
\end{equation}
while the BCES bisector fit\footnote{All our BCES fits used 1000 realizations of the relevant datasets.} provides $\alpha=0.444\pm0.060$ and $\beta=2.34\pm0.077$.
The \texttt{FITEXY} method, adapted to include intrinsic scatter \cite[following][]{Tremaine2002_Msigma}, provides
$\alpha=0.167\pm0.029$ and $\beta=2.004\pm0.038$ 
(and an intrinsic scatter of 0.24) -- in excellent agreement with the BCES(Y|X) result.
Fig.~\ref{fig:gam_tot_lledd_all} presents the BCES bisector and FITEXY best-fitting relations.
Table~\ref{tab:corr_fits} lists the best-fitting parameters for all three linear fits, as well for the other statistically significant $\gamx-\lledd$ correlations we find for the primary BASS sample of \Ngood\ sources (i.e., those with $P<0.1\%$).
We also tabulate the standard deviation of the residuals (i.e., $\sigma[\gamx^{\rm obs}-\gamx^{\rm fit}]$).
We note that in fitting this $\gamx-\lledd$ relation, as well as in virtually all other cases,  the best-fitting BCES bisector linear regression diverges from the two other methods, and the resulting residuals show significant trends with \lledd.

We highlight the fact that the best-fitting $\gamx-\lledd$ relations we find using the consistent BCES(Y|X) and FITEXY(Y|X) methods suggest much weaker dependence of \gamx\ on \lledd, compared to those reported in previous studies (i.e., $\alpha\simeq0.16$ vs.\ $\sim0.31$). 
Moreover, these linear relations fail to reduce the considerable amount of scatter in the $\gamtot-\lledd$ plane: the standard deviations of the residuals, roughly $\sigma(\Delta)\gtrsim0.25$, are comparable to the general standard deviation of \gamtot\ in our sample ($\sigma[\gamtot]=0.27$).
Thus,
there is little evidence that these linear relations provide a preferred  description of the $\gamtot-\lledd$ parameter space, and/or the range in \gamtot\ seen in the BASS sample.

Despite the statistically significant (though weak) $\gamtot-\lledd$ correlation found for the primary BASS sample as a whole, a closer inspection of the three different \mbh\ subsets provides very limited evidence for such correlations within these subsets.
In particular, the subsets with ``direct'', ``single-epoch'', and ``\sigs'' determinations of \mbh\ result in $P$-values of 94.6, 0.36, and 39.2\%, respectively (all based on Spearman correlation tests; see Table~\ref{tab:corr_pars}).
We highlight the lack of a statistically significant correlation among the most reliable \mbh\ determinations (i.e., the ``direct'' subset) and the weak evidence for a correlation among the single-epoch subset, which most closely resembles the \mbh\ estimation methodology of the aforementioned reference studies.

These apparently qualitatively inconsistent results -- for the BASS sample as a whole and for the different \mbh\ subsets -- suggest that the correlation between \gamtot\ and \lledd\ may not be robust nor universal.
We discuss this further in Section~\ref{sec:discuss}.

\begin{figure*}
\includegraphics[width=0.485\textwidth]{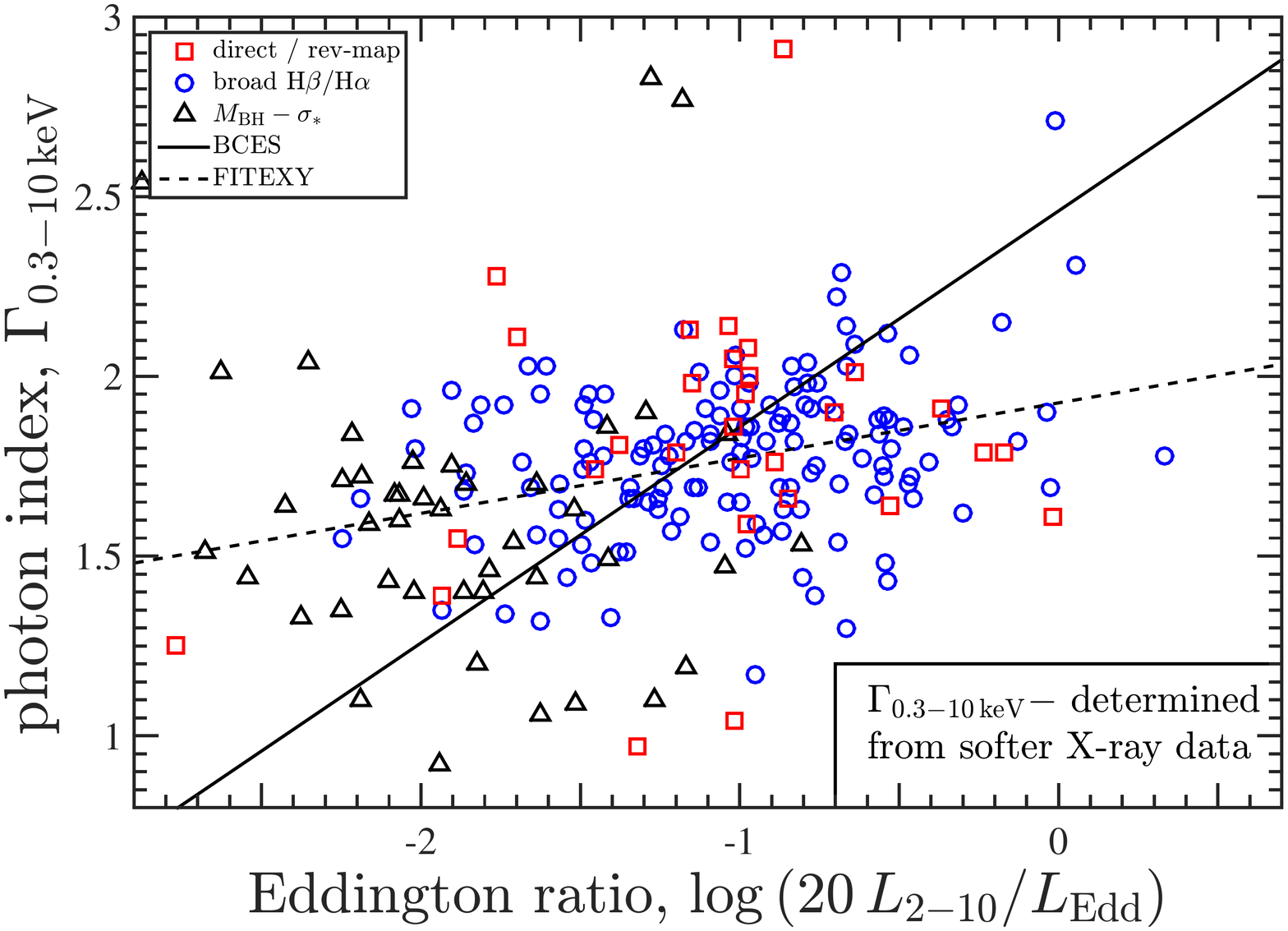}
\hfill
\includegraphics[width=0.485\textwidth]{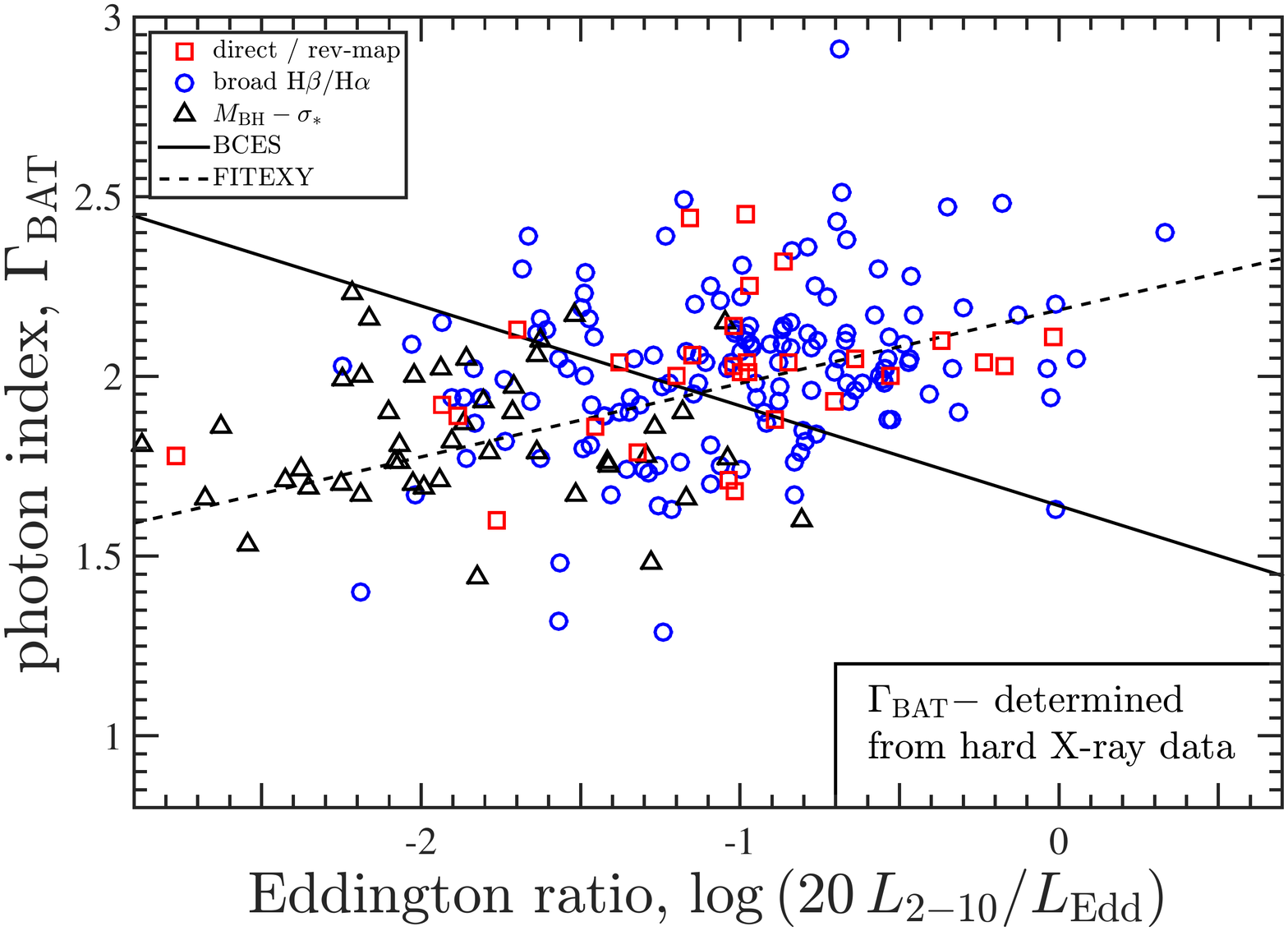}\\
\includegraphics[width=0.485\textwidth]{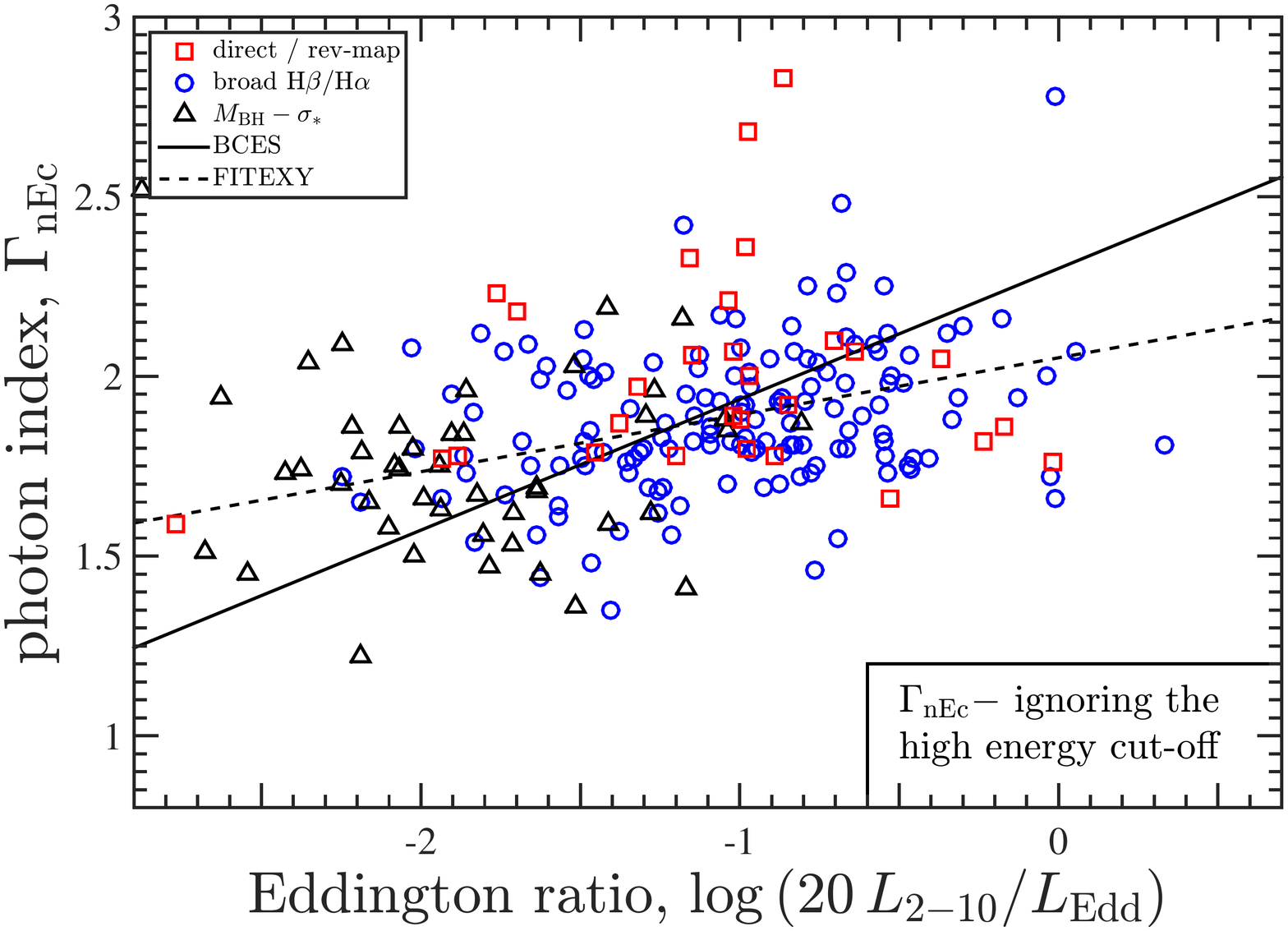}
\caption{
Same as Fig.~\ref{fig:gam_tot_lledd_all}, but with alternative determination of the photon index, \gamx.
{\it Top-Left}: \gamsoft\ -- obtained from the full X-ray spectral model, fitted over the energy range 0.3-10\,\kev.
{\it Top-Right}: \gambat\ -- obtained from a power-law spectral model fitted over the \swift/BAT energy range of 14-195\,\kev.
{\it Bottom}: \gamnec\ -- obtained from a modified spectral model that ignores the high-energy exponential cut-off.
}
\label{fig:gam_lledd_other_gam}
\end{figure*}

\subsection{Examining alternative determinations of \gamx\ and/or \lledd}
\label{subsec:res_other_gam}

\begin{figure*}
\includegraphics[width=0.485\textwidth]{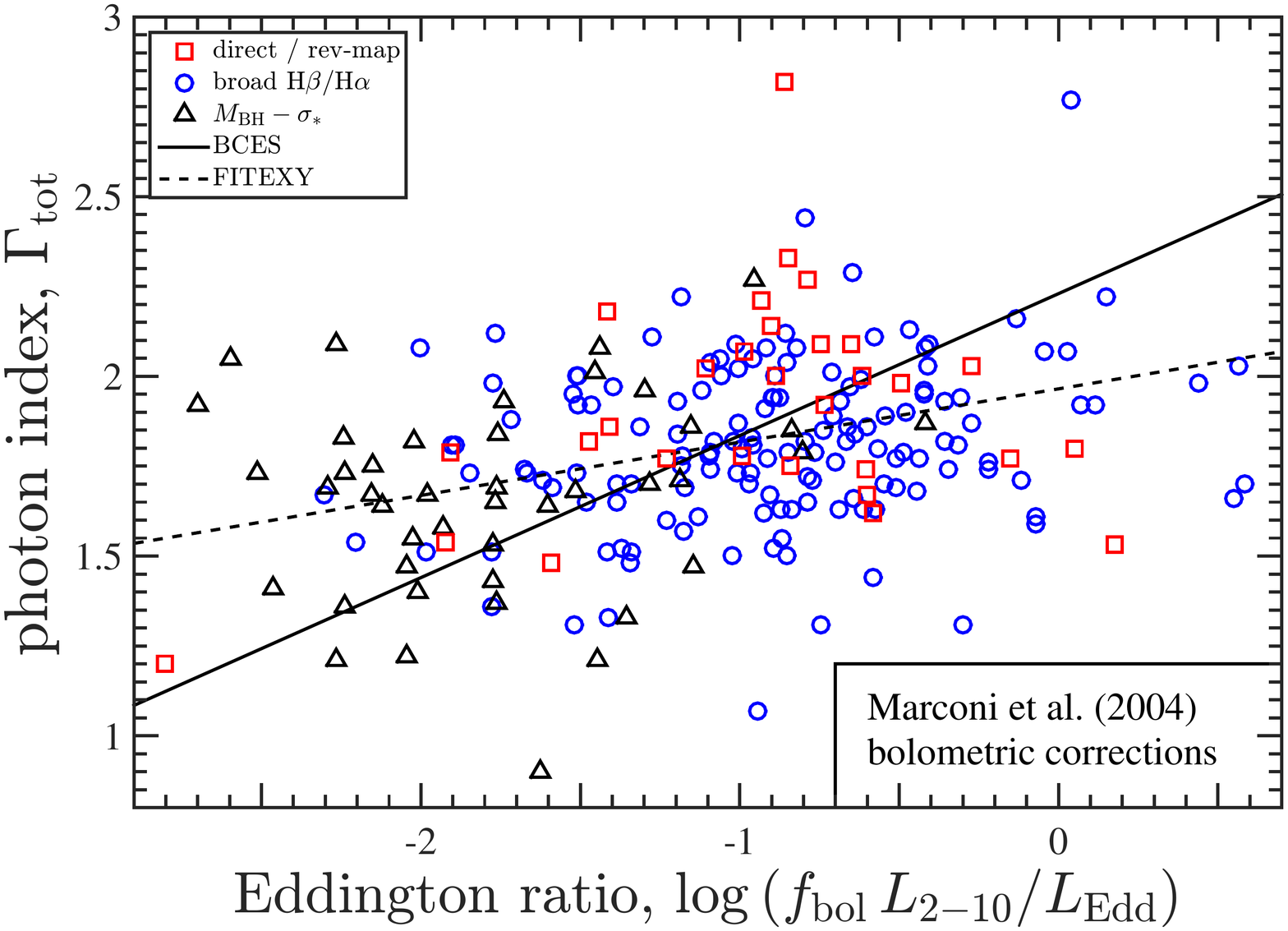}
\hfill
\includegraphics[width=0.485\textwidth]{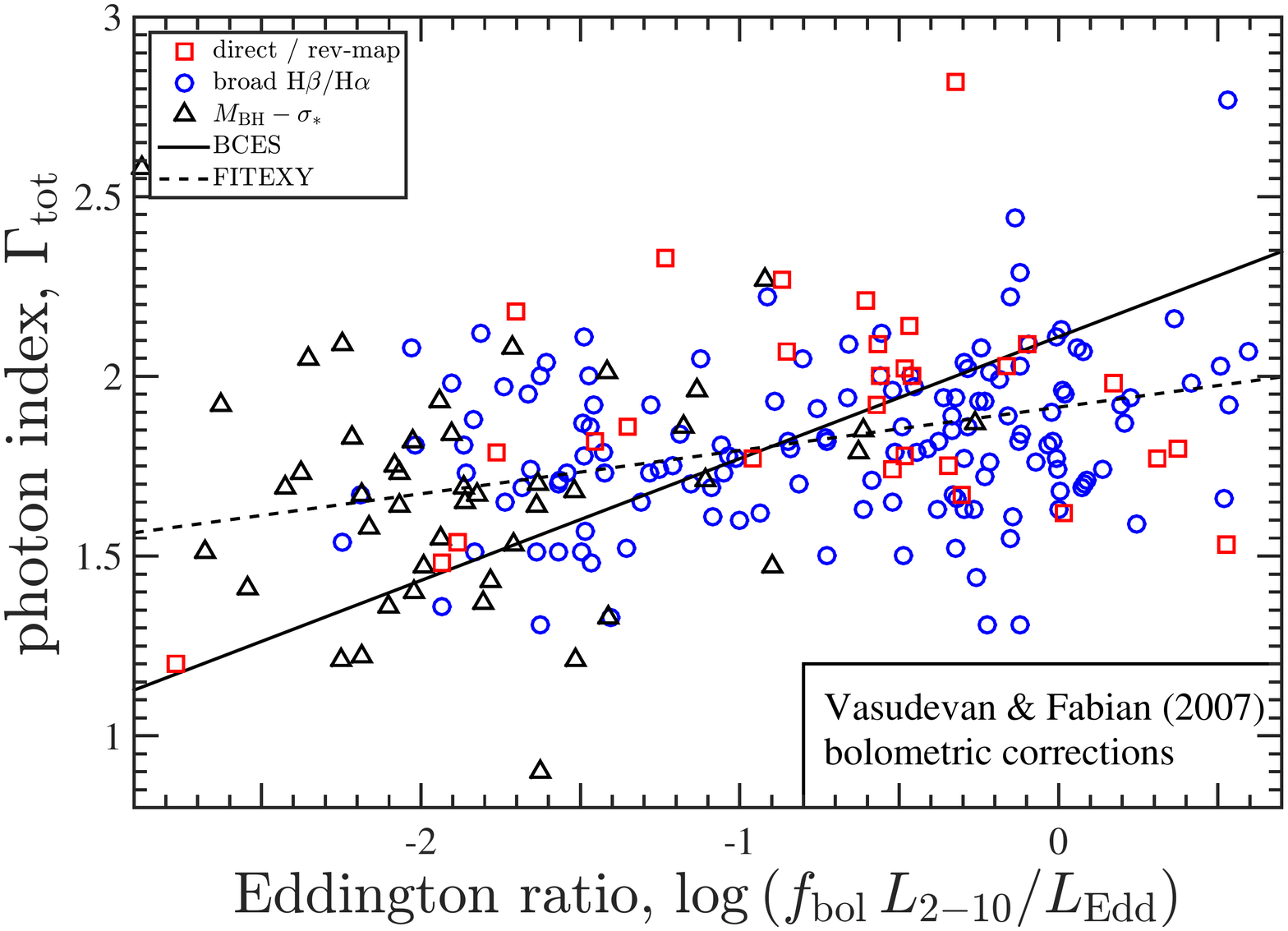}\\
\includegraphics[width=0.485\textwidth]{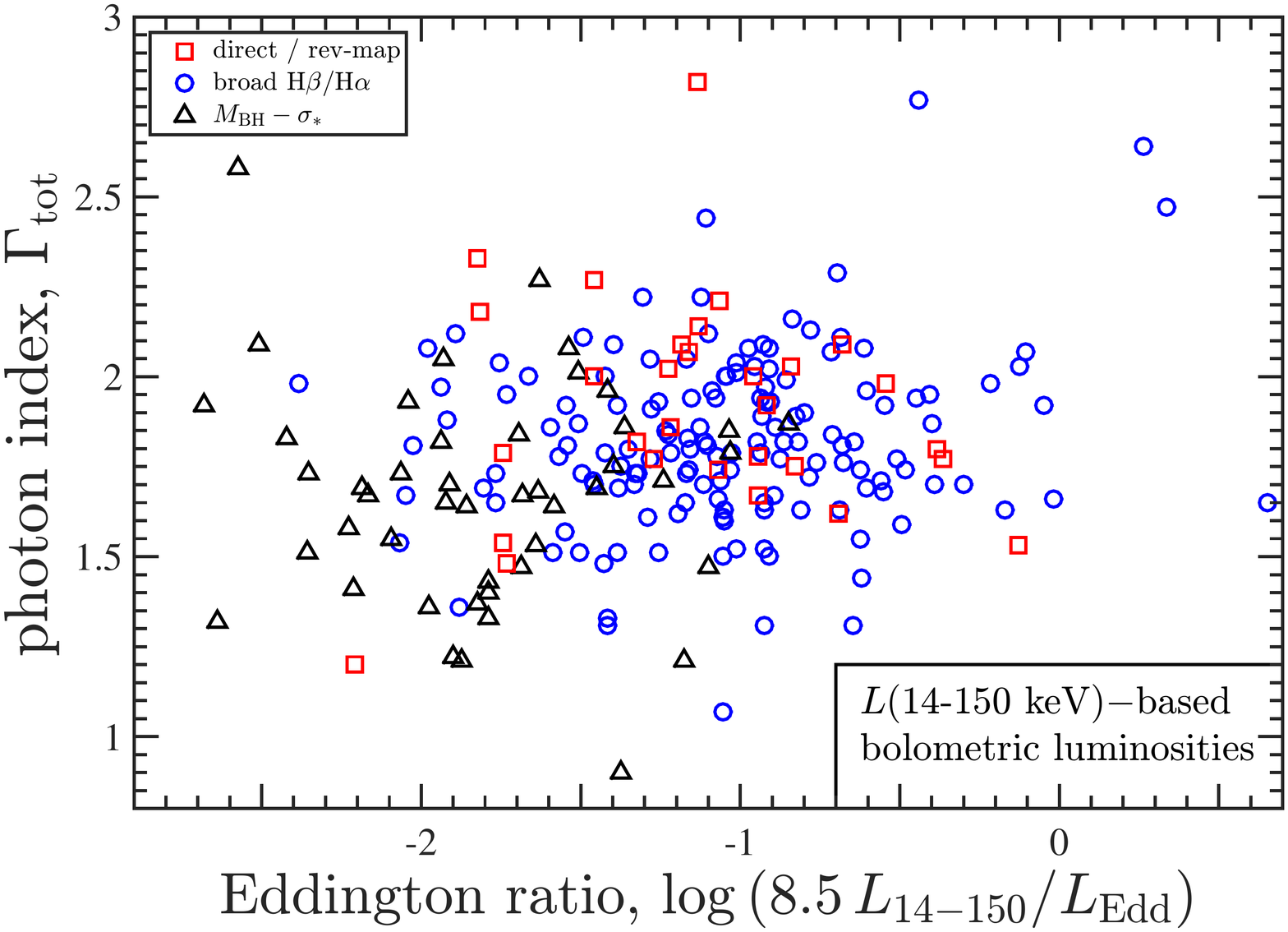}
\hfill
\includegraphics[width=0.485\textwidth]{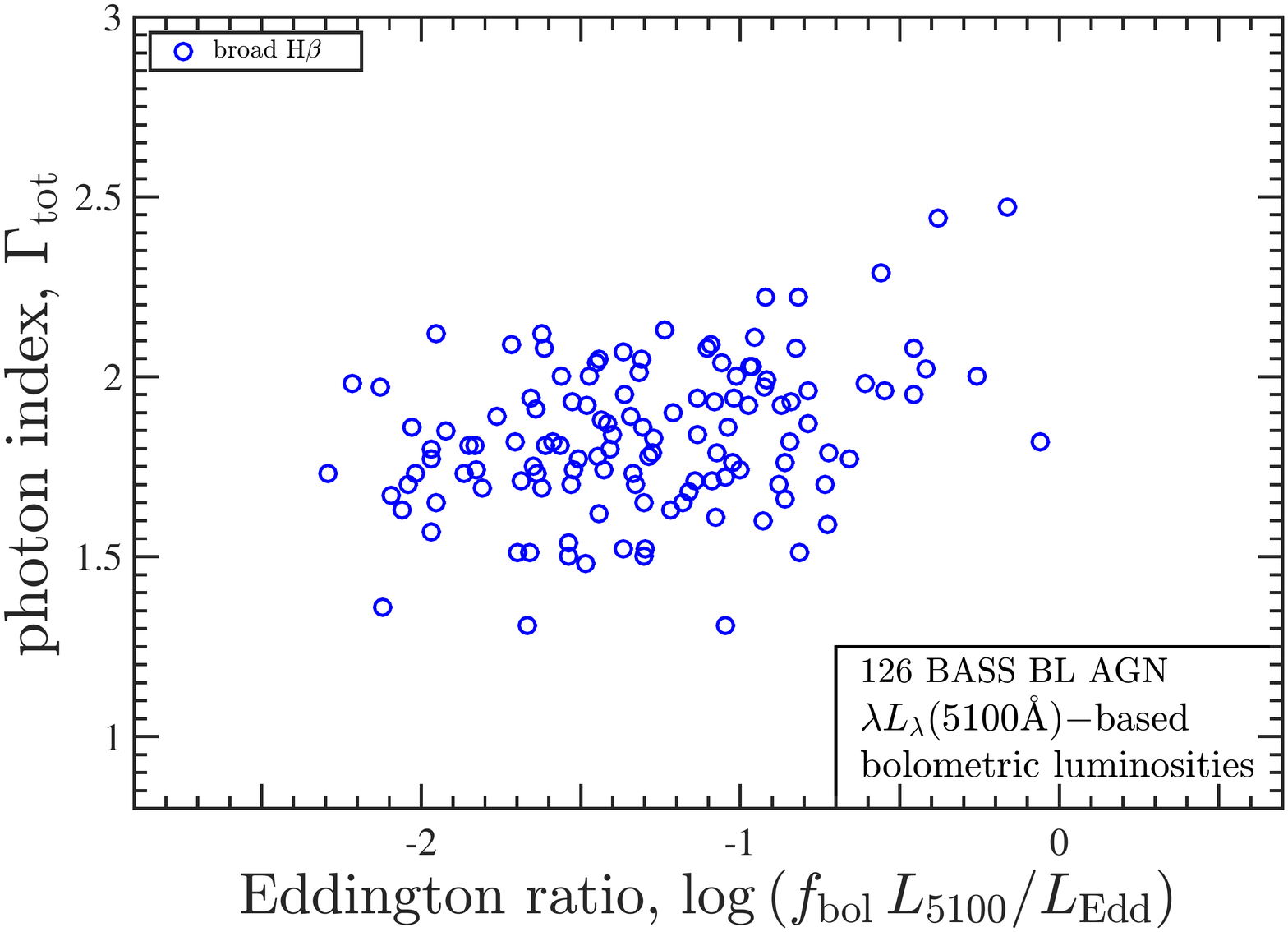}\\

\caption{
Same as Fig.~\ref{fig:gam_tot_lledd_all}, but with alternative determination of the accretion rate, \lledd.
The symbols in all panels are as in Figs.~\ref{fig:gam_tot_lledd_all}-\ref{fig:gam_lledd_other_gam}. 
Black diagonal lines represent the best-fitting correlations (using either the BCES bisector or the FITEXY methods), for the cases where the $\gamx-\lledd$ correlation is statistically significant.
{\it Top}: \gamtot\ vs. \lledd\ estimates derived from \Lhard, and through either the \Lhard-dependent bolometric corrections of \citet[][{\it left}]{Marconi2004}, or the \lledd-dependent bolometric corrections of \citet[][{\it right}]{VasudevanFabian2007_BC}.
{\it Bottom-Left}: \gamtot\ vs. \lledd\ estimates derived through $\Lbol=8.5\times\Lbat$. 
{\it Bottom-Right}: \gamtot\ vs. \lledd\ estimates derived from \Lop, and through \Lop-dependent bolometric corrections (calibrated against those of \citealt{Marconi2004}; see Section~\ref{subsec:Lbol_lledd}), for the subset of \Nmbhhb\ AGN with single-epoch, broad \hbeta\ determinations of \mbh.
In these two cases, we find no statistically significant correlations.
}
\label{fig:gam_lledd_other_lledd}
\end{figure*}

We next examine the alternative determinations of \gamx\ and \lledd\ available for our sample, to further test whether we can establish any (stronger) relations between these quantities.
In particular, we have examined relations between the \Lhard-based estimates of \lledd\ (and $\fbolhard=20$), and either \gamsoft, \gambat, or \gamnec\ -- shown in Fig.~\ref{fig:gam_lledd_other_gam}.
We have also used the alternative set of \lledd\ estimates, in which \Lbol\ is estimated from \Lhard\ and the bolometric corrections of either \cite{Marconi2004}, or those of \cite{VasudevanFabian2007_BC} -- presented in the top two panels of Fig.~\ref{fig:gam_lledd_other_lledd}.
The \Lbat-based estimates of \lledd\ (i.e., $\Lbol=8.5\,\Lbat$) are presented in the bottom-left panel of Fig.~\ref{fig:gam_lledd_other_lledd}, 
while the bottom-right panel presents the \Lop-based estimates of \lledd, for the subset of \Nmbhhb\ AGN for which \mbh\ is determined from single-epoch spectroscopy of the broad \hbeta\ line.

The results of all these tests are qualitatively similar to our main analysis of \gamtot\ vs. \lledd(\Lhard): large scatter, statistically significant correlations between \gamx\ and \lledd\ for the overall primary BASS sample (i.e., \Ngood\ AGN), but no correlation within any of the three \mbh\ subsets --
as can be seen in the results of the formal correlation analyses (listed in Table~\ref{tab:corr_pars}).
We particularly note that in all the cases we examined (i.e., all \lledd), the most reliable ``direct'' \mbh\ subset did \emph{not} result in statistically significant correlations.
The ``single-epoch'' subset shows somewhat stronger evidence for correlations, with $P$-values $\lesssim 1\%$ in all cases, and a statistically significant (but weak) correlation for the case where \gamnec\ is considered ($P\simeq10^{-2}\,\%$, $r_{\rm s}=0.31$; the best-fit (FITEXY) relation has $\alpha=0.16$).
Another noteworthy exception is the lack of correlation between \gamtot\ and the \Lbat-based determinations of \lledd, even among the entire primary BASS sample (bottom panel of Fig.~\ref{fig:gam_tot_lledd_all}).
Importantly, we find that the correlation between \gamtot\ and the \Lop-based estimates of \lledd, for the subset of AGN with single-epoch, broad \hbeta\ determinations of \mbh, is neither truly statistically significant ($P=0.11\%$) nor strong ($r_{\rm s}=0.29$). 
We will revisit this subset when comparing our results with previous studies of the $\gamx-\lledd$ relation (see Section~\ref{subsec:comparison}).

We finally note that the BASS sample provides no compelling evidence for an ``inversion'' of the $\gamx-\lledd$ relation for low-\lledd\ systems (i.e., changing into an anti-correlation for $\lledd \ll 0.01$), as suggested by some studies \cite[e.g.,][]{Younes2011_LINERS,Kamizasa2012_IMBHs_X_var,Yang2015_Gamma_LLedd,Kawamuro2016_lowL_BAT_Suzaku}.
Motivated by these claims, we also explicitly verified that the high-\lledd\ regime in our sample (AGN with $\lledd>0.01$) does not present a strong, underlying $\gamx-\lledd$ correlation which is then ``diluted'' by the lower-\lledd\ sources (see Section \ref{subsec:res_sanity_checks} below and Appendix~\ref{app:sanity_checks}).

\subsection{Additional tests for subsets of AGN}
\label{subsec:res_sanity_checks}

Finally, we examined several subsets of sources within our BASS sample, verifying that none of the choices we made in defining our sample, or our treatment of certain physically-motivated spectral components, would have a significant effect on our conclusion.
In particular, we tested for the existence of $\gamx-\lledd$ correlations among: 
AGN with $0.05<z<0.5$ - minimizing aperture effects;
AGN with high-quality (SDSS) optical spectra;
AGN with $0.01<\lledd<1$; 
AGN with no heavy obscuration ($\log(\NH/\cmii)<23$); 
and
AGN without warm absorbers.
These subsets are described in Appendix~\ref{app:sanity_checks}, and the results of the correlation tests are tabulated in Table~\ref{tab:corr_pars_minor}).
The qualitative results of this analysis are consistent with what we find for the primary BASS sample: for each subset, we find either no correlation, or alternatively, a weak correlation for all the AGN in that subset, while finding no correlations among sources with differing \mbh\ determination methods.

\subsection{Relations between \gamx\ and other AGN properties}
\label{subsec:res_other_props}

\begin{figure}
\includegraphics[width=0.475\textwidth]{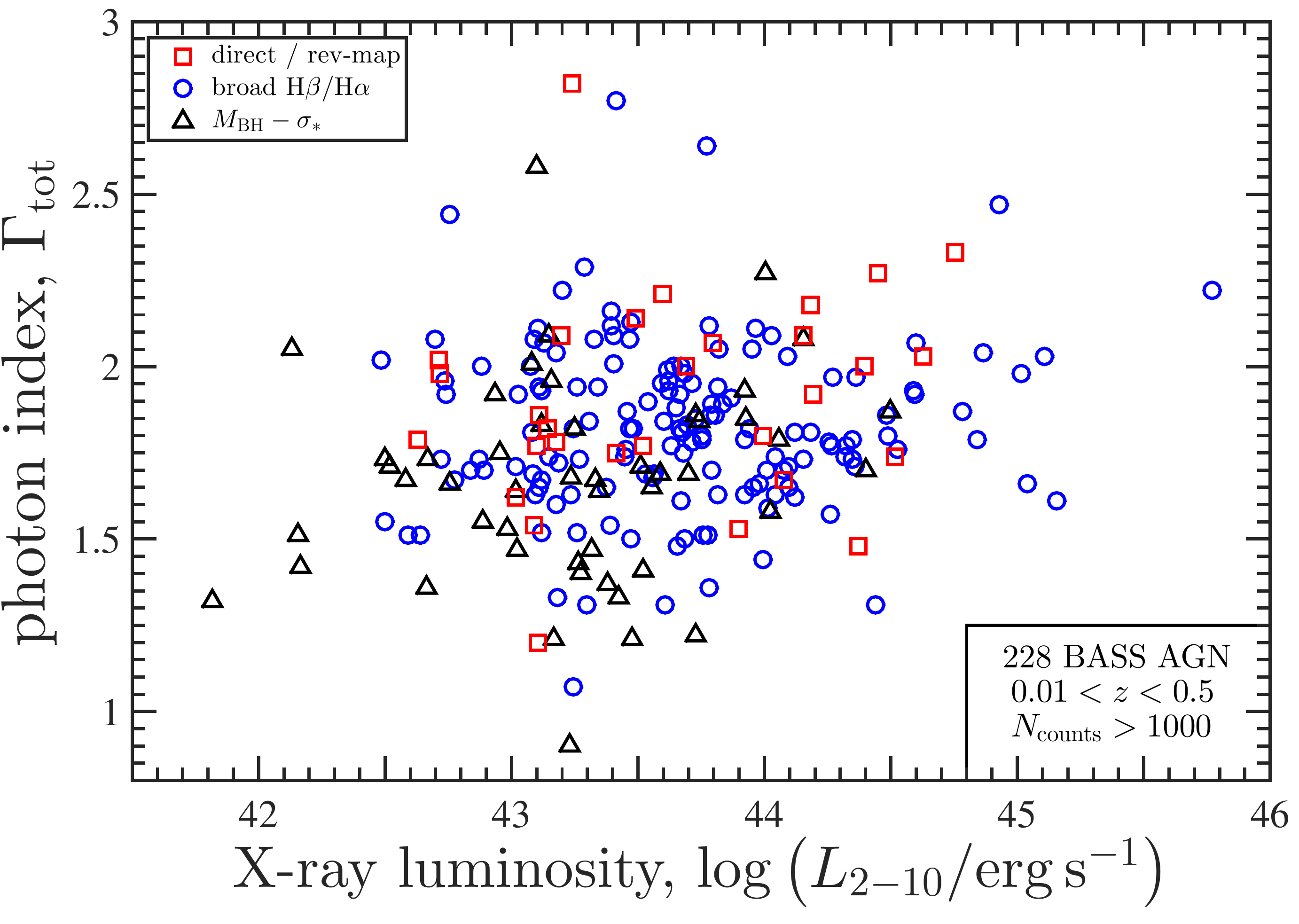}\\
\includegraphics[width=0.475\textwidth]{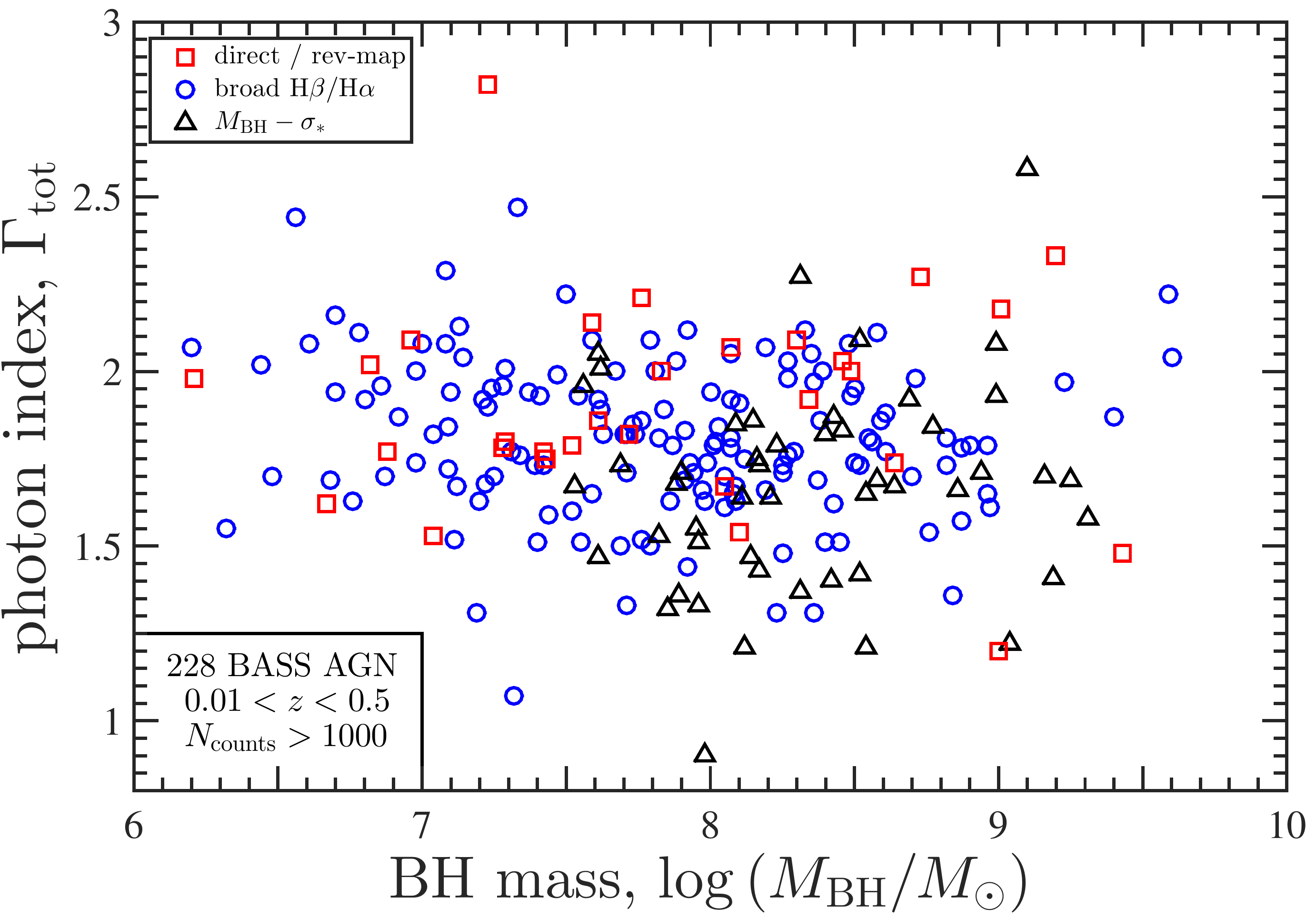}\\
\includegraphics[width=0.475\textwidth]{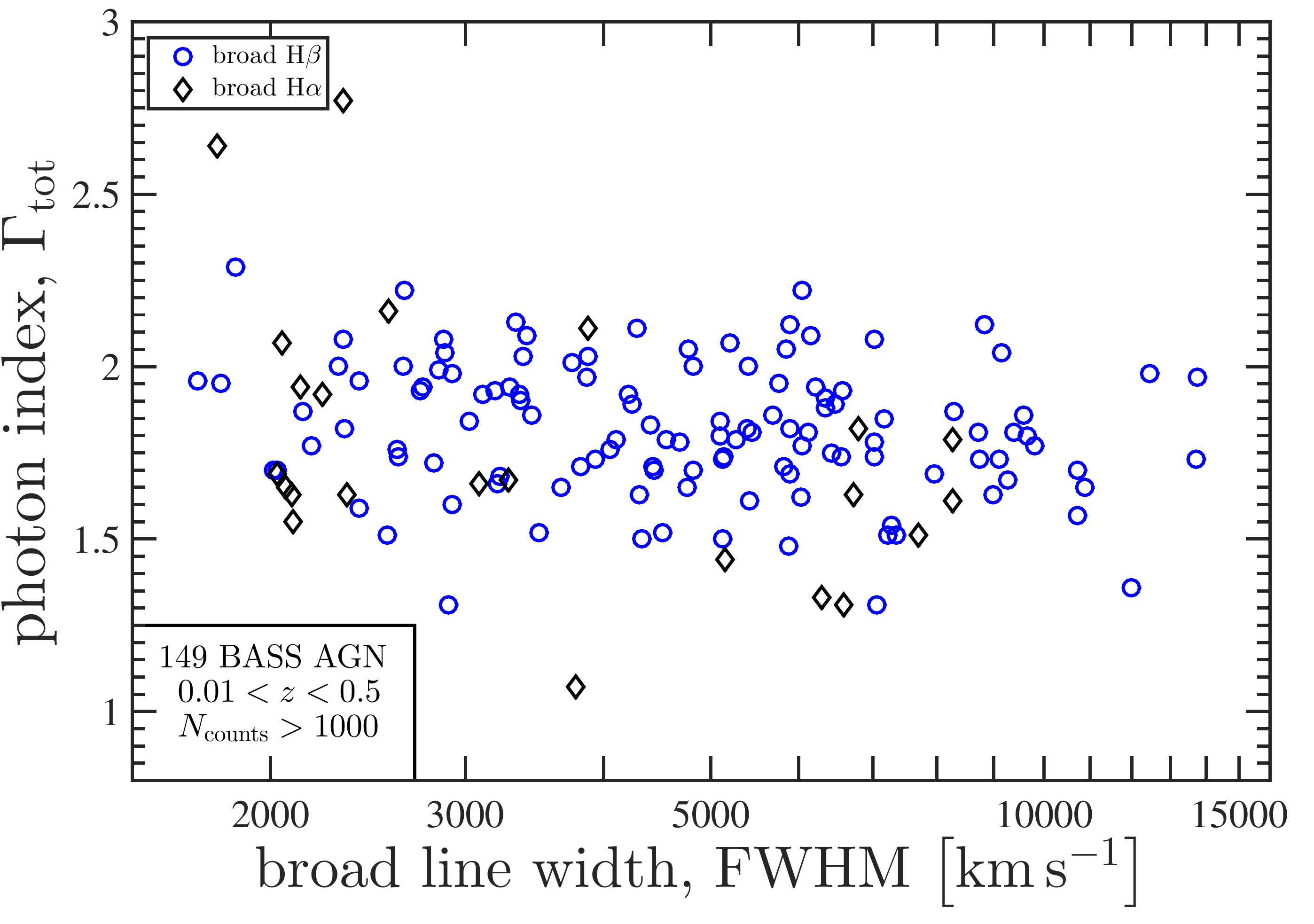}\\
\caption{
Testing for relations between X-ray photon index \gamtot\ and other key AGN properties. 
{\it Top}: \gamtot\ vs.\ hard X-ray luminosity (14-150\,\kev) probed by \swift/BAT, \Lbat.
{\it Centre}: \gamtot\ vs.\ BH mass, \mbh.
In both these panels, symbols are identical to Fig.~\ref{fig:gam_tot_lledd_all}.
{\it Bottom}: \gamtot\ vs.\ width of the broad Balmer lines (\Hbeta\ or \Halpha) for those sources for which these data are available.
No correlations are found between \gamtot\ and any of these properties.
}
\label{fig:gam_tot_other_props}
\end{figure}

We looked for relations between \gamx\ and other key properties of the accreting SMBHs in our sample.
Fig.~\ref{fig:gam_tot_other_props} presents \gamtot\ vs.\ \Lhard, \fwhb\ (or \ha), and \mbh.\footnote{For the purposes of the test with the FWHM of broad Balmer lines, we focused only on the \Nmbhse\ AGN with single-epoch determinations of \mbh\ (i.e., ignoring the \Nmbhdirect\ AGN with ``direct'' mass measurements).}
The $P$-values associated with these correlation tests are listed in Table~\ref{tab:corr_pars}.
None of these relations resulted in a statistically significant correlation. 
A qualitatively similar result was obtained when testing for correlations involving \gamsoft\ or \gamnec.
These results are in agreement with the findings of previous studies that investigated possible links between \gamx\ and other AGN properties.

The broad dynamical range in  \Lhard, \Lbol, \mbh, and \lledd\ covered by our sample allows us to 
further investigate whether the mutual dependence between (some of) these quantities has any effects on the  $\gamx-\lledd$. 
To this end, we examined subsets of our sample for which one of these properties is controlled.
Considering only the AGN with $\log(\Lhard/\ergs)=43.25-43.75$ (i.e., a bin of $\pm0.25$ dex around the median luminosity, with 71 sources), we find no evidence for a significant $\gamx-\lledd$ correlation ($P=2\%$ for \gamtot, and $>0.1\%$ for all other cases).
This should be compared to the highly significant correlations found when considering the entire luminosity range ($P\ll 10^{-3}\%$ in all cases; see Table~\ref{tab:corr_pars}).
A similar analysis for AGN with $\log\left(\mbh/\Msun\right)=7.75-8.25$ (again within $\pm0.25$ dex of the median value; 64 sources) provides a qualitatively different result: 
the statistically significant $\gamx-\lledd$ correlation holds for most cases ($P\simeq0.1\%$ for \gamtot\ and \gamsoft; $< 10^{-3}\%$ for \gamnec\ and \gambat).
We stress that these two special subsets of AGN cover the same range in both \lledd\ and \gamx\ as does our primary BASS sample. 
This is only possible thanks to the broad range of \lledd\ and \mbh\ provided through the BASS project (see K17).

These results, together with the fact that \lledd\ is strongly correlated with \Lhard\ in our sample ($P\simeq10^{-3}\%$, $r_{\rm s}=0.29$),
suggest that the $\gamtot-\lledd$ relation for the primary BASS sample may be -- at least partially -- driven by the trend with source luminosity.


\begin{table*}
\centering
\caption{BASS $\gamx-\lledd$ correlations: significance tests}
\label{tab:corr_pars}
\begin{tabular}{lclcrc} 
\hline
\Lbol\ & $\Gamma$ & sub-sample & $N$ &  $P-{\rm value}$ & $r_{\rm s}$ \\ 
tracer & ~~~      & ~~~        &     &  [\%]            & ~~~         \\
\hline
$C\cdot\Lhard$ & \gamtot     & all, $0.01<z<0.5$  & \Ngood      & $\mathbf{1.7\times10^{-4}\%}$ & 0.311  \\
~~~~~~~~~~~~~  & ~~~~~~~~    & direct \mbh\       & \Nmbhdirect &      94.6\%       & 0.013 \\
~~~~~~~~~~~~~  & ~~~~~~~~    & single-epoch \mbh\ & \Nmbhse     &       0.36\%      & 0.237  \\
~~~~~~~~~~~~~  & ~~~~~~~~    & \sigs-based  \mbh\ & \Nmbhsigs   &      39.2\%       & 0.125  \\
\hline
$C\cdot\Lhard$ & \gamsoft    & all, $0.01<z<0.5$  & \Ngood      & $\mathbf{1.2\times10^{-4}\%}$ & 0.315 \\ 
~~~~~~~~~~~~~  & ~~~~~~~~    & direct \mbh\       & \Nmbhdirect &      66.8\%                   & 0.082 \\
~~~~~~~~~~~~~  & ~~~~~~~~    & single-epoch \mbh\ & \Nmbhse     &       1.31\%                  & 0.203  \\
~~~~~~~~~~~~~  & ~~~~~~~~    & \sigs-based  \mbh\ & \Nmbhsigs   &      67.4\%                   &-0.062 \\
\hline
$C\cdot\Lhard$ & \gamnec     & all, $0.01<z<0.5$  & \Ngood      & $\mathbf{3.9\times10^{-7}\%}$ & 0.377 \\ 
~~~~~~~~~~~~~  & ~~~~~~~~    & direct \mbh\       & \Nmbhdirect &      78.5\%                   & 0.052 \\
~~~~~~~~~~~~~  & ~~~~~~~~    & single-epoch \mbh\ & \Nmbhse     & $\mathbf{1.1\times10^{-2}\%}$ & 0.311 \\
~~~~~~~~~~~~~  & ~~~~~~~~    & \sigs-based  \mbh\ & \Nmbhsigs   &      40.7\%                   & 0.121 \\
\hline
$C\cdot\Lhard$ & \gambat     & all, $0.01<z<0.5$  & \Ngood      & $\mathbf{2.0\times10^{-8}\%}$ & 0.405 \\
~~~~~~~~~~~~~  & ~~~~~~~~    & direct \mbh\       & \Nmbhdirect &       4.00\%                  & 0.377 \\
~~~~~~~~~~~~~  & ~~~~~~~~    & single-epoch \mbh\ & \Nmbhse     &       0.16\%                  & 0.256 \\
~~~~~~~~~~~~~  & ~~~~~~~~    & \sigs-based  \mbh\ & \Nmbhsigs   &      40.8\%                   & 0.121 \\
\hline
$\fbol[{\rm M04}]\cdot\Lhard$  & \gamtot & all, $0.01<z<0.5$  & \Ngood      & $\mathbf{8.7\times10^{-5}\%}$ & 0.319 \\
~~~~~~~~~~~~~                  & ~~~~~~~ & direct \mbh\       & \Nmbhdirect & 84.2\% & 0.038 \\
$\fbol[{\rm VF07}]\cdot\Lhard$ & \gamtot & all, $0.01<z<0.5$  & \Ngood      & $\mathbf{1.6\times10^{-4}\%}$ & 0.311 \\ 
~~~~~~~~~~~~~                  & ~~~~~~~ & direct \mbh\       & \Nmbhdirect & 94.6\% & 0.013 \\
$C\cdot\Lbat$                  & \gamtot & all, $0.01<z<0.5$  & \Ngood     & 0.39\% &  0.190 \\
~~~~~~~~~~~~~                  & ~~~~~~~ & direct \mbh\       & \Nmbhdirect & 48.4\% & -0.133 \\
$\fbol[{\rm M04}]\cdot\Lop$    & \gamtot & single-epoch, \hbeta & \Nmbhhb   & 0.11\% &  0.287 \\
\hline
\hline
AGN property & $\Gamma$ & sub-sample & $N$ &  $P-{\rm value}$ & $r_{\rm s}$ \\
\hline
\Lhard & \gamtot & all, $0.01<z<0.5$   & \Ngood\    &  2.43\% &  0.149 \\
\Lbat\ & ~~~~~~~ & ~~~~~~~~~~~~~~      & ~~~~~~~    & 85.3\%  & -0.012 \\
\mbh\  & ~~~~~~~ & ~~~~~~~~~~~~~~      & ~~~~~~~    &  3.27\% & -0.142 \\
FWHM(\hb/\ha) &  & ~~~~~~~~~~~~~~      & \Nmbhse\   & 84.3\%  & -0.013 \\
\hline
\end{tabular}
\end{table*}

\begin{table*}
\centering
\caption{BASS $\gamx-\lledd$ correlations: best-fit parameters}
\label{tab:corr_fits}
\begin{tabular}{lccccccccccc} 
\hline
\Lbol\ & $\Gamma$ & 
\multicolumn{3}{c}{BCES (bisector)}  & 
\multicolumn{3}{c}{BCES (Y|X)}  & 
\multicolumn{4}{c}{FITEXY} \\
tracer & ~~~    & 
$\alpha$ &  $\beta$ & $\sigma(\Delta)$ & 
$\alpha$ &  $\beta$ & $\sigma(\Delta)$ & 
$\alpha$ &  $\beta$ & $\epsilon$ & $\sigma(\Delta)$ \\
\hline
$C\cdot\Lhard$  & \gamtot       & 
$0.444\pm0.060$ & $2.34\pm0.07$ & 0.33 &
$0.167\pm0.040$ & $2.00\pm0.05$ & 0.26 &
$0.167\pm0.029$ & $2.00\pm0.04$ & 0.24 & 0.26 \\
~~~~~~~~~~~~~~  & \gamsoft      &  
$0.601\pm0.096$ & $2.46\pm0.12$ & 0.45 & 
$0.130\pm0.050$ & $1.89\pm0.07$ & 0.30 &
$0.154\pm0.029$ & $1.93\pm0.04$ & 0.24 & 0.30 \\
~~~~~~~~~~~~~~ & \gamnec     & 
$0.364\pm0.150$ & $2.30\pm0.18$ & 0.30 & 
$0.159\pm0.031$ & $2.06\pm0.18$ & 0.26 & 
$0.158\pm0.029$ & $2.05\pm0.04$ & 0.24 & 0.26 \\
~~~~~~~~~~~~~~ & \gambat     & 
$-0.278\pm0.070$ & $1.64\pm0.09$ & 0.35 &
$ 0.160\pm0.023$ & $2.17\pm0.03$ & 0.21 & 
$ 0.204\pm0.018$ & $2.18\pm0.03$ & 0.04 & 0.22 \\
\hline
$\fbol[{\rm M04}]\cdot\Lhard$ & \gamtot & 
$0.395\pm0.060$  & $2.23\pm0.07$ & 0.33 &
$0.143\pm0.032$ & $1.96\pm0.04$ & 0.26 &
$0.148\pm0.023$ & $1.97\pm0.03$ & 0.22 & 0.26 \\
$\fbol[{\rm VF07}]\cdot\Lhard$ & \gamtot & 
$0.339\pm0.048$ & $2.11\pm0.05$ & 0.33 &
$0.116\pm0.026$ & $1.91\pm0.03$ & 0.26 & 
$0.120\pm0.019$ & $1.91\pm0.02$ & 0.22 & 0.26 \\
\hline
\end{tabular}
\end{table*}

\section{Discussion}
\label{sec:discuss}

\subsection{The BASS $\gamx-\lledd$ plane for different classes of AGN}
\label{subsec:robustness}

Our analysis shows no evidence for a robust $\gamx-\lledd$ relation among the subsets of AGN for which reliable estimates of \mbh\ (and therefore, of \lledd) are available, while \emph{also} showing evidence for a significant correlation among the BASS sample as a whole, as well as (marginal) evidence for a correlation among the broad-line sources.
How could these qualitatively contradicting results be reconciled?

The study of \cite{Winter2012_X_SEDs_WAs} has identified a similar discrepancy, when finding a strong $\gamx-\lledd$ correlation \emph{only} among the broad-line \swift/BAT selected AGN in their sample. 
The interpretation put forward by that study suggested that the lower-luminosity and/or lower-\lledd, absorbed AGN are found in a different accretion state.
For our BASS sample, a closer inspection of Figs.~\ref{fig:gamma_stats} and \ref{fig:gam_tot_lledd_all}, suggests that the \sigs\ subset (i.e., AGN with no broad Balmer lines, and no direct \mbh\ determination) exhibits somewhat lower \gamx, compared with the other two \mbh\ subsets \cite[see also][]{Vasudevan2013_XBR_SED}.
In addition, the studies of \cite{Fabian2008_rad_press} and \cite{Fabian2009_rad_press_Swift} showed that such narrow-line sources are predominantly low-\lledd\ systems. 
The combined effect of these two trends is that the \sigs\ subset mainly extends towards the low-\lledd, low-\gamx\ part of the parameter space, which in turn results in statistically significant $\gamx-\lledd$ correlations once this subset is included in the analysis.

Are these two trends driven by physical processes or by observational limitations (i.e., selection effects)?
As suggested by \cite{Fabian2008_rad_press}, the tendency of obscured (narrow line) AGN towards low \lledd\ is likely driven by the limited radiation pressure that low-\lledd\ AGN exert on the surrounding dusty circumnuclear gas (i.e., the dusty tori), which in turn result in increased levels of optical and X-ray obscuration.

The outstanding question is therefore whether the somewhat lower \gamx\ seen in obscured AGN is \emph{driven} by the $\gamx-\lledd$ correlation (the origin of which not yet well understood; see below), or rather by an unrelated physical and/or observational effect, which -- when combined with the tendency of obscured AGN to have lower \lledd\ -- \emph{produces} the observed $\gamx-\lledd$ relation for our entire sample of BASS sources. 
One such scenario would be if obscured AGN have multiple partially-covering (i.e., clumpy) absorption components \cite[e.g.,][]{Cappi1996_NGC5252_X}.
In such a case, the measured \gamx\ might be flatter than the real underlying photon index.
Thus, a scenario in which the lower \gamx\ of obscured sources is driven by physical effects beyond the $\gamx-\lledd$ correlation would require that many (or indeed, most) obscured sources would have (at least) two partially covering absorbing components. 
This is, arguably, a rather extreme scenario.
%

\begin{figure}
\includegraphics[width=0.475\textwidth]{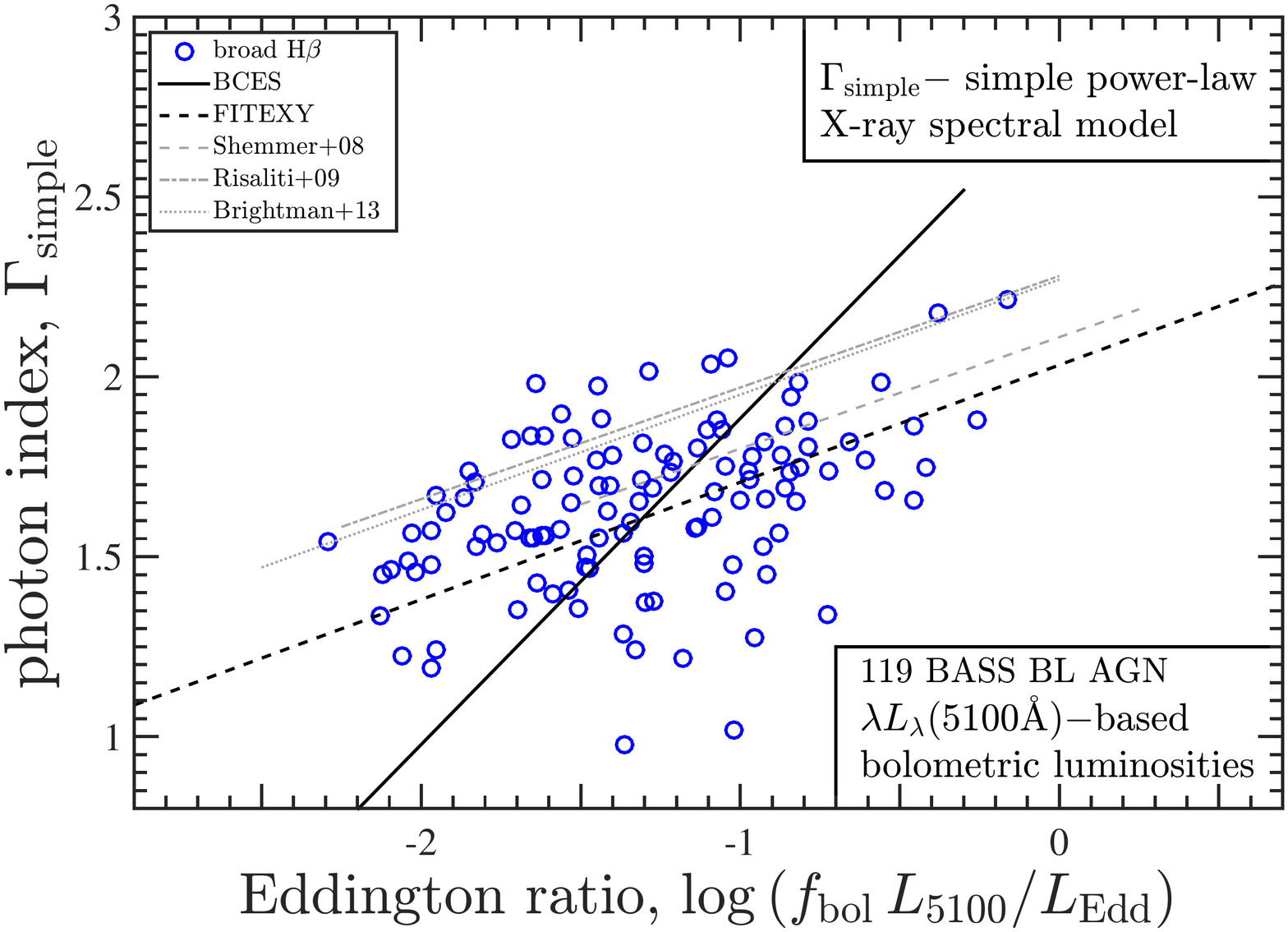}\\
\includegraphics[width=0.475\textwidth]{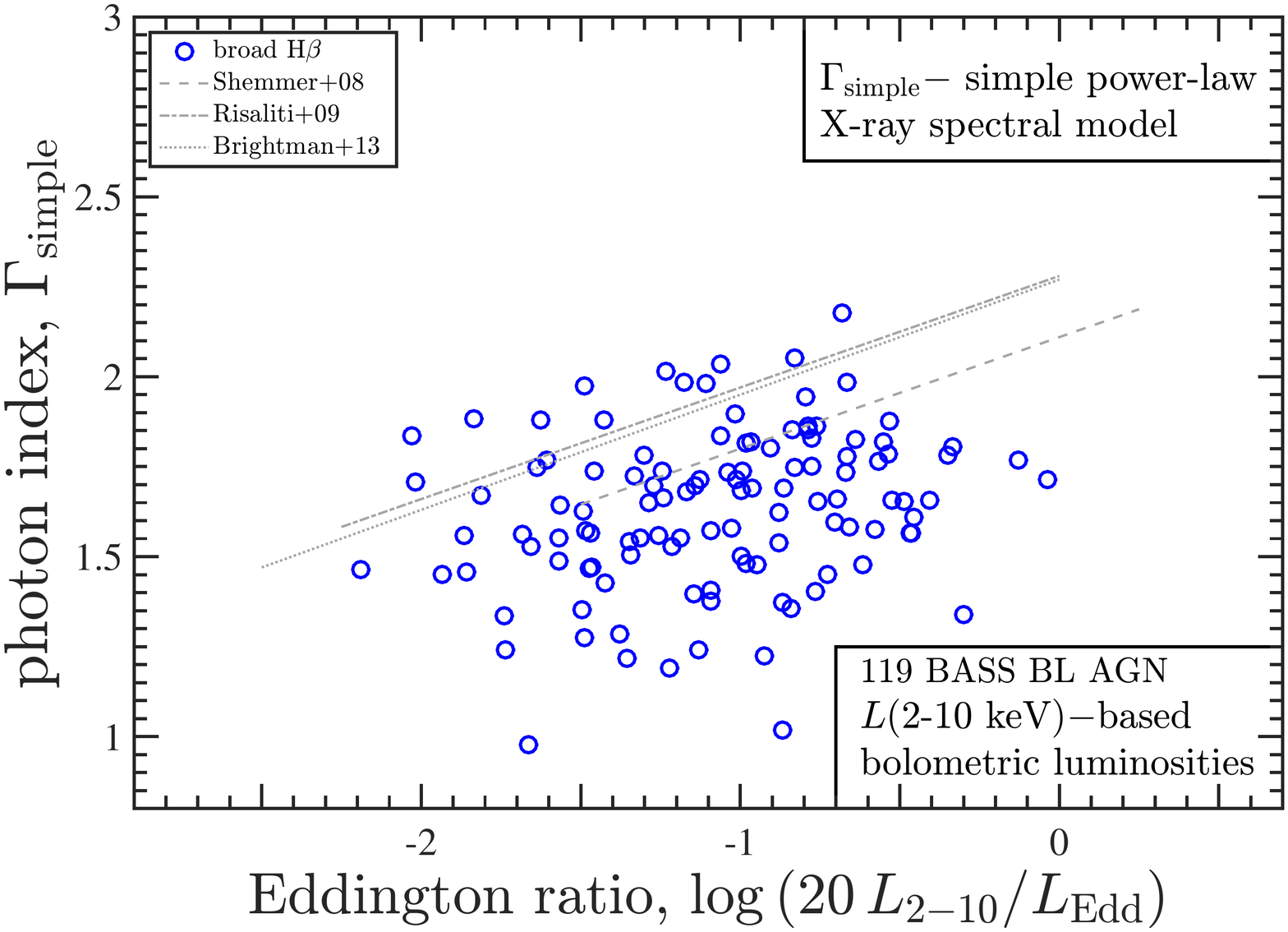}
\caption{
An attempt to compare BASS with previous studies of the $\gamx-\lledd$ relation in broad-line, $z\gtrsim1$ AGN. 
Here, the photon index \gamsim\ is derived from a simplified spectral model of a power-law, fit to the softer X-ray data of a subset of 119 unobscured ($\log(\NH/\cmii)\le22$), broad-line AGN where \mbh\ is determined from broad \hbeta.
{\it Top}: \lledd\ is estimated from \Lop. 
These data exhibit a statistically significant correlation, unlike what we found when considering \gamtot\ (c.f.\ the bottom-right panel of Fig.~\ref{fig:gam_lledd_other_lledd}).
{\it Bottom}: \lledd\ is estimated from \Lhard. No correlation is found in this case.
}
\label{fig:gam_simple_lledd}
\end{figure}

\subsection{Comparison with previous studies}
\label{subsec:comparison}

We demonstrated that our sample of \Ngood\ low-redshift, hard X-ray selected AGN shows no significant evidence for a correlation between the hard X-ray photon index, \gamx, and the normalized accretion rate, \lledd, nor with other key AGN properties such as BH mass (\mbh) and/or hard X-ray luminosity (\Lbat).
This stands in contrast to the findings of several studies.
In what follows, we briefly summarize three such studies, which form the main reference for our comparison.
\begin{itemize}

\item 
\citet[S08]{Shemmer2008_Gamma_LLedd} studied 35 high-luminosity, high-redshift quasars (at $z\sim0-3.5$), for which the 
X-ray spectral analysis mostly relied on \xmm\ data in the observed-frame energy range 0.5-10\,\kev. 
BH masses were determined from broad \hbeta\ spectroscopy, using the same prescription we use here, and \lledd\ were calculated through the \Lop-dependent prescription of \cite{Marconi2004}, consistent with the \Lop-dependent bolometric corrections we use here.

\item
\citet[R09]{Risaliti2009_Gamma_LLedd} analysed a sample of 343 moderate-to-high luminosity ($43 \ltsim \log[\Lx/\ergs] \ltsim 46.7$) SDSS quasars at $0.1 \ltsim z \ltsim 4.5$, with archival \xmm\ data \cite[compiled by][]{Young2009_SDSS_XMM_DR5}.
The X-ray spectra, covering 0.5-10\,\kev, were fitted with an (absorbed) power-law model.
BH masses were determined from either the \hbeta, \MgII, or \CIV\ broad emission lines (with 314 AGN having the more reliable \hbeta- or \mgii-based masses).
Bolometric luminosities were derived by using a fixed-shape UV-optical SED, and a power-law X-ray SED (with $E_{\rm C}=100\,\kev$).
The $\gamx-\lledd$ relations found for the subsets of AGN with either \hbeta- or \mgii-based \mbh\ determinations are markedly different ($\alpha=0.58$ and $0.24$, respectively).

\item
\citet[B13]{Brightman2013_Gamma_Ledd} analysed a sample of 69 X-ray selected, broad-line AGN from the \chandra\ surveys in the E-CDF-S and COSMOS fields, covering $0.5\ltsim z \ltsim2$ and $42.5 \ltsim \log[\Lx/\ergs] \ltsim 45.5$.
The sample was restricted to sources with more than 250 counts in their spectra.
BH masses were obtained through either \halpha- or \mgii-based single-epoch estimators, which are generally consistent with those used here and in the other reference studies.

\end{itemize}
All these studies, which serve as primary reference studies for our work, focused on unobscured, broad-line AGN, for which \mbh\ is determined through single-epoch spectroscopy of broad emission lines -- comparable to our ``single-epoch'' \mbh\ subset. 
In addition, most of these studies employed a spectral model that includes only a single power-law, with a minor absorption correction for a few sources.

The study of \cite{Winter2012_X_SEDs_WAs} employed a more elaborate X-ray spectral model to a sample of broad-line \swift/BAT-selected AGN, and identified strong $\gamx-\Lx$ and $\gamx-\lledd$ relations, although the slope of the latter ($\alpha=0.23$) is somewhat flatter than what is found for the optically-selected quasars mentioned above.
More recently, the study of \cite{Brightman2016_Gamma_masers} studied the $\gamx-\lledd$ relation in a sample of nine heavily obscured (mostly Compton-thick) AGN, for which precise \mbh\ measurements are available from resolved megamaser kinematics. This analysis resulted in a significant correlation with best-fit parameters that are consistent with those derived in the aforementioned studies of unobscured AGN.

Comparing these reference studies to our BASS analysis, 
we first note the higher quality and broader energy coverage of our BASS X-ray data.
These allow for a much more elaborate and robust spectral decomposition, taking into account several physically-motivated components, and provide a set of various determinations of the key quantities (i.e., \gamx\ and \Lbol).
We also note that our sample completely overlaps with the reference studies in terms of the range of \gamx\ and \lledd\ covered, and that it includes \Ntypeone\ broad-line AGN -- the only class of AGN studied in the reference studies. 

Although at face value our BASS analysis suggests a $\gamx-\lledd$ correlation which is similar to those found in the reference studies, we note two main differences.
First, we stress that we find little evidence for any $\gamx-\lledd$ link among BASS sources for which \mbh\ is determined from single-epoch spectra of broad emission lines -- the only subset comparable with the reference studies. 
Even for this subset, the only statistically significant correlation we find is when using \gamnec\ (which may be similar to the \gamx\ used in some of the reference studies).
Moreover, the correlation involving \gambat\ -- which could be thought of as comparable to what is measured for high-redshift sources (see S08) -- is insignificant (although at $P=0.16\%$).
Second, the slopes of the best-fitting relations we derive for our \emph{entire} BASS sample (Table~\ref{tab:corr_fits}) differ from those previously reported ($\alpha\simeq0.3$): we find $\alpha\simeq0.16$ for the (\gamx|\lledd) correlation analyses, but $\alpha\gtrsim0.4$ for the BCES bisector. 
The discrepancy between the different fitting methods probably reflects the large scatter in the $\gamx-\lledd$ plane.

To allow for a more direct comparison, 
we have derived yet another set of \gamx\ measurements which aims to resemble the analysis performed in previous studies.
We re-fitted the X-ray data of 162 BASS AGN that have $\log(\NH/\cmii)\le 22$ with a simplified spectral model of an absorbed power-law over the rest-frame energy range $2-10\,\kev$. 
By ignoring any additional components (i.e., warm absorbers, reflection, Fe K$\alpha$), this model -- and the chosen energy range -- are similar to what was used in the aforementioned reference studies.
We stress that these derived photon indices, \gamsim, are \emph{not} identical to \gamsoft\ (see Section~\ref{subsec:xray_data}), despite the similarity in the respective energy ranges, as \gamsoft\ was derived from a more elaborate spectral model.
We further focus on those AGN for which \mbh\ is determined through single-epoch spectroscopy of the broad \hbeta\ emission line, and on the \Lop-based estimates of \lledd.

Figure~\ref{fig:gam_simple_lledd} (top panel) plots these simplified photon indices (\gamsim) against the \Lop-based estimates of \lledd\ for the relevant 119 AGN in our sample.
In this case, we find a statistically significant ($P\simeq2\times10^{-3}\%$) yet, again, weak ($r_{\rm s}=0.383$) correlation between these two particular quantities.
We recall that a similar analysis, with \Lop-based estimates of \lledd\ for the single-epoch, broad-\hbeta\ subset, but with \gamtot, yielded only a marginally significant correlation ($P=0.1\%$; see Section~\ref{subsec:res_other_gam} and Fig.~\ref{fig:gam_lledd_other_lledd}).
A formal correlation analysis results in 
$(\alpha,\beta)=(0.906\pm0.11,2.79\pm0.13)$, 
$(0.304\pm0.09,2.00\pm0.13)$, and 
$(0.326\pm0.08,2.03\pm0.10)$, 
for the BCES bisector, BCES(Y|X), and FITEXY methods, respectively (with an intrinsic scatter of 0.2 added in the latter case).
The best-fit slopes of the latter two (Y|X) relations are in excellent agreement with those reported by the main three reference studies.
We stress that we find \emph{no} significant correlation between \gamsim\ and the primary, \Lhard-based estimates of \lledd\ ($P=0.8\%$), as seen in the bottom panel of Fig.~\ref{fig:gam_simple_lledd}.
This is an important point, as some of the studies that reported strong $\gamx-\lledd$ correlations  (e.g., R09, B13) relied, at least partially, on \Lhard-based determinations of \lledd, and not on purely \Lop-based ones.
%

\begin{figure*}
\includegraphics[width=0.475\textwidth]{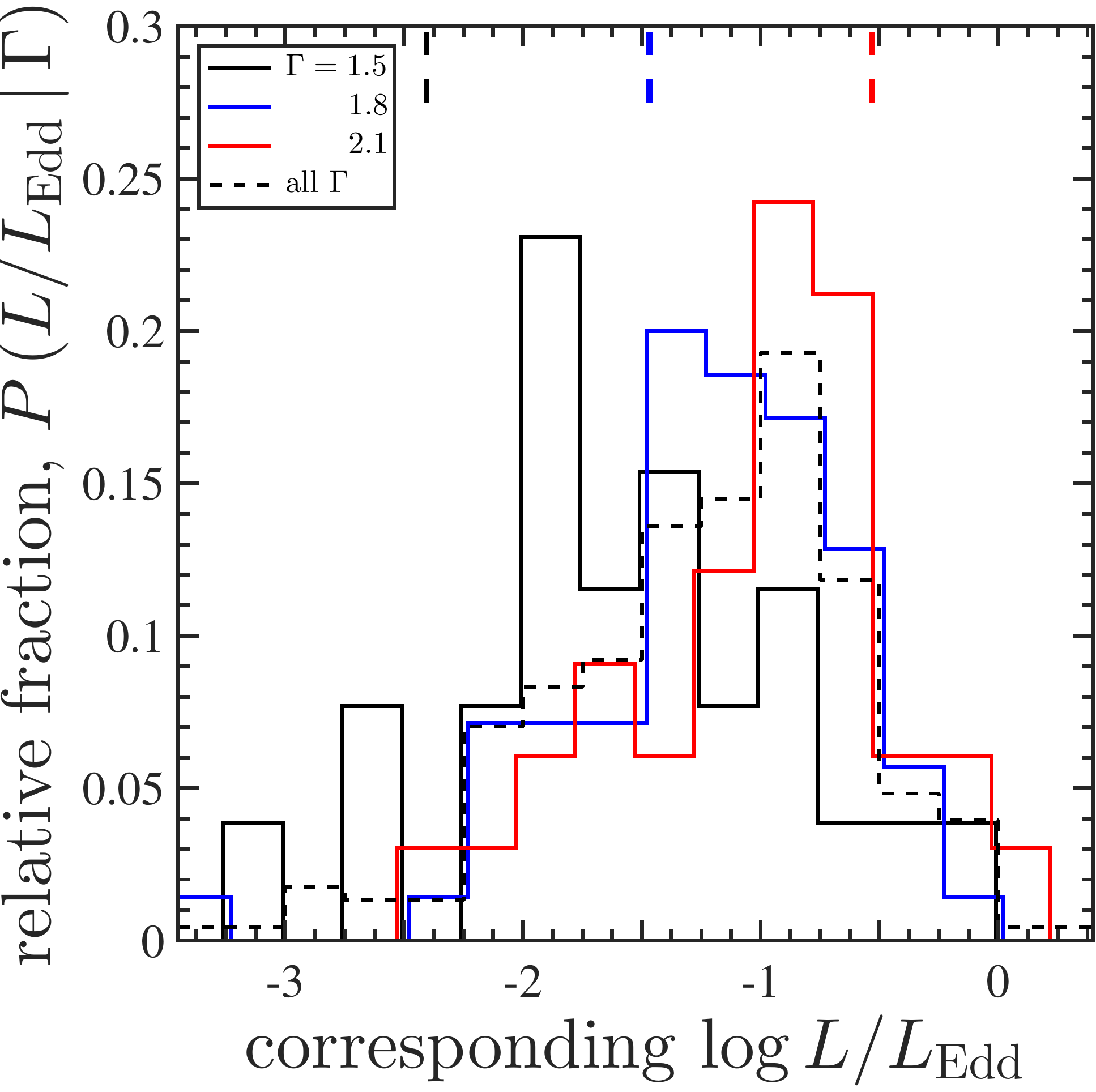}
\hfill
\includegraphics[width=0.475\textwidth]{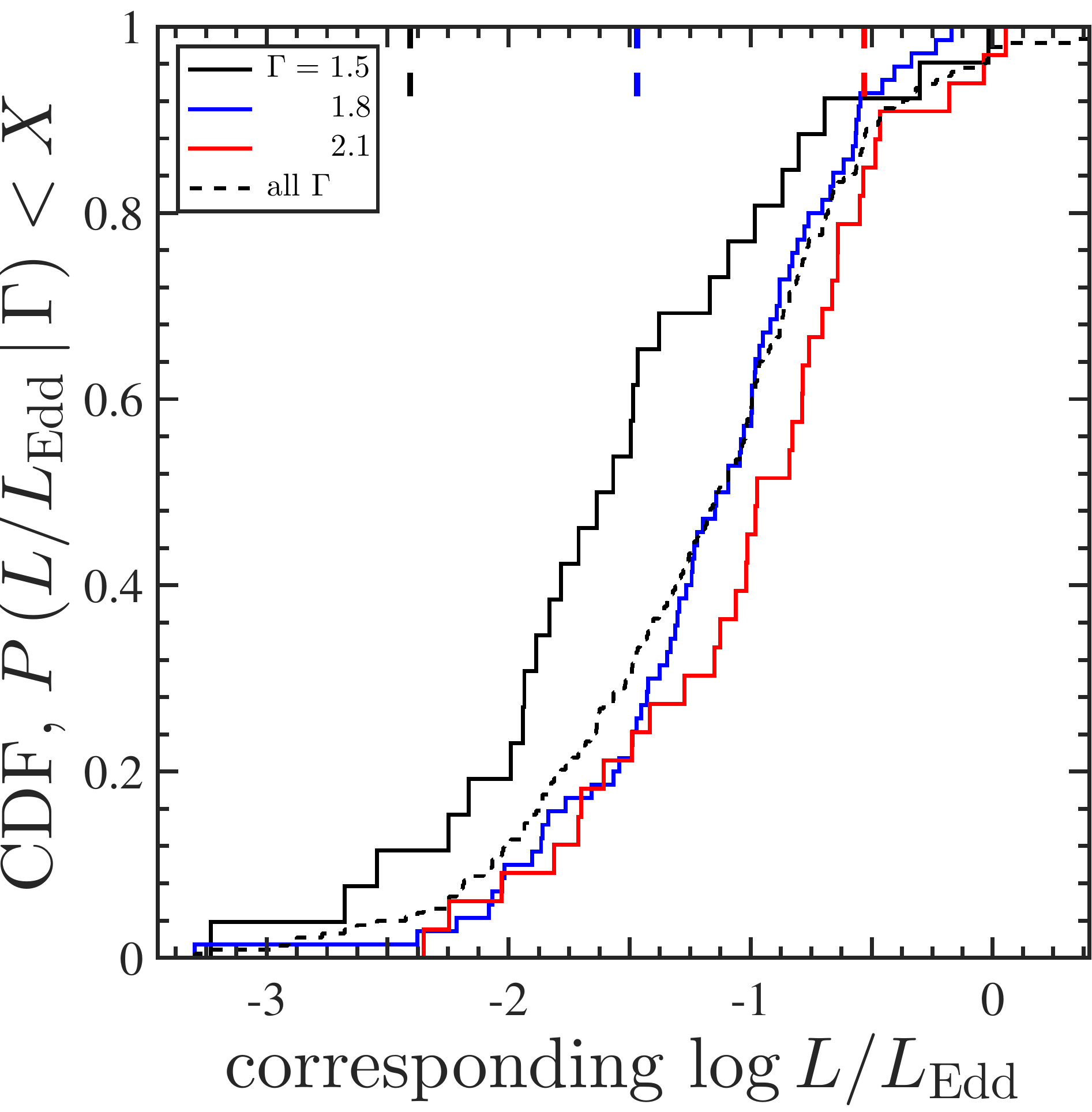}
\caption{
Testing the usability of \gamx\ as a predictor of \lledd.
{\it Left}: the \emph{observed} distribution of \lledd\ for the \Ngood\ BASS sources in our sample, split into three non-overlapping bins of \gamtot: $\gamtot=1.6$, $1.8$ and $2.1$ (all bins have widths $\pm0.1$; solid lines). 
The dashed histogram traces the distribution of \lledd\ among the entire sample of \Ngood\ BASS sources.
The short vertical dashed lines near the top mark the \lledd\ predicted from the $\gamx-\lledd$ relation reported in \citet{Brightman2013_Gamma_Ledd}.
{\it Right}: same distributions of \lledd, but represented as cumulative fractions. 
Both panels demonstrate the significant scatter in \lledd\ at nearly fixed \gamx; the significant overlap between the range of \lledd\ covered by each sub-sample; and the differences between the distributions' peaks (or medians) and the ``predicted'' values.
}
\label{fig:lledd_pred}
\end{figure*}

%
Thus, it appears that the photon index derived from a simplified X-ray spectral model of a power-law traces a $\gamx-\lledd$ correlation that is very similar to what was reported in previous studies.
However, such a correlation is \emph{not} seen when using more elaborate X-ray spectral models, nor when using X-ray-based determinations of \lledd.
We conclude that the tension between our overall result for the BASS AGN -- of no strong $\gamx-\lledd$ relation for broad-line AGN -- and that of previous studies, might be indeed driven by the limited spectral coverage, or the simplified spectral model used in some of the reference studies.

\subsection{Using \gamx\ as a BH growth indicator}
\label{subsec:pred_lledd}


As first pointed out by \cite{Shemmer2008_Gamma_LLedd},
one of the exciting implications of a strong and tight relation between \gamx\ and \lledd\ is the possibility to use large X-ray surveys to construct nearly complete distributions of \lledd, particularly for high-redshift sources in deep extragalactic fields, where this key quantity is otherwise hard to measure \cite[e.g,][]{TrakhtNetzer2012_Mg2,Trakhtenbrot2016_COSMOSFIRE_MBH}.
The study of \cite{Brightman2013_Gamma_Ledd} can be considered as a demonstration of such an approach within a dedicated survey (COSMOS).
Moreover, the recent study of \cite{Brightman2016_Gamma_masers} suggested that this approach may also be applicable to heavily obscured (Compton-thick) AGN, potentially providing a unique probe of the accretion rates among these elusive objects.

However, our sample and analysis highlight the limitations associated with using \gamx\ measurements to predict \lledd.
We first recall that the overall scatter in the $\gamx-\lledd$ plane is large ($\sim$0.3 dex; see Figs.~\ref{fig:gam_tot_lledd_raw}--\ref{fig:gam_lledd_other_lledd}), and that the few statistically significant relations we find between \gamx\ and \lledd\ are weak (i.e., have flat slopes, $\alpha\simeq0.15$).

To further assess the usability of \gamx\ as a predictor of \lledd, we show in Fig.~\ref{fig:lledd_pred} the distributions of \lledd\ among three subsets of BASS sources with (almost) fixed \gamx\ values, $\gamx=1.5$, $1.8$, and $2.1$, with the \gamx\ bins defined within $\pm0.1$ of these values.
Here we used \gamtot\ and the \Lhard-based estimates of \lledd, for the sample of \Ngood\ BASS sources at $0.01<z<0.5$ with high-quality X-ray data ($\Ncnts>1000$) -- the same measurements as those presented in Fig.~\ref{fig:gam_tot_lledd_all}.
For all three \gamx\ bins, the corresponding distributions of \lledd\ span about two orders of magnitude, covering the entire range $\lledd\sim0.01-1$. 
This range is comparable to what is observed for other large samples of luminous AGN, at least out to $z\sim2$ 
\cite[e.g.,][and references therein]{TrakhtNetzer2012_Mg2,KellyShen2013_BHMF,Schulze2015_BHMF}.
For the $\gamx=1.8\pm0.1$ bin alone,\footnote{Corresponding to the median value of \gamtot\ for our primary sample (see Fig.~\ref{fig:gamma_stats}).} the 1-$\sigma$ interquartile range in \lledd\ (i.e., corresponding to the $16-84\%$ quantile range) covers roughly 1.1 dex. 
This is qualitatively similar to the behaviour of the overall distribution of \lledd\ in our sample (i.e., regardless of \gamx; dashed lines in Fig.~\ref{fig:lledd_pred}).
Moreover, the ranges in \lledd\ for the three \gamx\ subsets show considerable overlap, and in particular the distributions of \lledd\ corresponding to the $\gamx=1.8$ bin is similar to that of the $\gamx=2.1$ bin
(a two-sided Kolmogorov-Smirnov test resulted in $P_{\rm K-S}=16.9\%$).
Finally, these distributions of \lledd\ do \emph{not} agree 
with the predictions of the $\gamx-\lledd$ relations, as demonstrated by the short vertical (dashed) lines plotted near the top of both panels in Fig.~\ref{fig:lledd_pred}, which mark the \lledd\ values predicted from the B13 $\gamx-\lledd$ relation, for each of our \gamx\ bins.

The limitations on measuring \lledd\ from \gamx\ are further demonstrated by the corresponding correlation analysis.
The best-fit BCES relation we find for our primary sample (i.e., \Ngood\ sources) is 
\begin{equation}
    \lledd = \left(0.71 \pm 0.27\right)\,\gamtot - \left(2.44\pm0.48\right) \,\, ,
	\label{eq:gamma_lledd_bces}
\end{equation}
which is consistent, within the considerable uncertainties, to the relation reported by S08 (their Eq.~2). 
The uncertainties on the best-fitting parameters in Eq.~\ref{eq:gamma_lledd_bces} are so large that for a given \gamtot\ with zero measurement uncertainty, they predict values of \lledd\ with a 1-$\sigma$ inter-percentile range of 1.37 dex (i.e., the 16-84\% percentile range).
Moreover, the corresponding FITEXY(\lledd\,|\,\gamtot) analysis suggests that a satisfactory fit, with $\chi^2/\nu\simeq1$, can only be obtained with the addition of a significant level of intrinsic scatter, exceeding 0.6.

Notwithstanding these limitations, it might still be possible to identify subsets of extremely high- or low-\lledd\ AGN, probed by correspondingly extreme \gamx\ (i.e., $\gamx\gtrsim2.3$ or $\lesssim 1.2$). 
This is supported by the relatively clear separation between the peaks (and medians) of the distributions seen for $\gamx=1.5$ and $2.1$ in Fig.~\ref{fig:lledd_pred}.
We note, however, that such extreme \gamx\ are only observed among a minority of AGN, out to $z\sim4$ \cite[e.g.,][]{Just2007_Xray_hiL,BrandtAlexander2015_Rev,Cappelluti2016_CANDELS_S,Marchesi2016_XVP_opt_ID}.

We conclude that the large scatter and weak correlations (at best) in the $\gamx-\lledd$ plane significantly hinder the prospects of using \gamx\ measurements to establish the distribution of \lledd\ among samples of high-redshift AGN.

\subsection{Possible physical links between \gamx\ and \lledd}
\label{subsec:physical_reasons}

Previous studies have tried to explain the positive $\gamx-\lledd$ correlation through a picture where the increasing \lledd\ is causing an increased UV radiation, which in turn causes more efficient cooling in the corona.
In principle, one may expect a similar trend of increasing \gamx\ with \emph{decreasing} \mbh, as in the framework of geometrically-thin, radiatively thick accretion discs this is also expected to increase the UV incident radiation \citep[e.g.,][]{DavisHubney2006,Done2012,DavisLaor2011_AD}.

We however recall that our analysis showed no correlation between \gamx\ and \mbh\ (Fig.~\ref{fig:gam_tot_other_props}).
One way to accommodate this lack of trend with the aforementioned physical picture is if the X-ray-emitting corona is located closer to the disk for lower-\mbh\ systems, therefore \emph{reducing} the amount of incident UV radiation. 
Such trends are indeed suggested by some reverberation mapping studies 
(see, e.g., \citealt{DeMarco2013_Xray_RM,Kara2013_Fe_RM}, and the review by \citealt{Uttley2014_Xray_RM_rev}).

We conclude that any scenario that connects the observed $\gamx-\lledd$ relation to variations in the UV radiation field that is up-scattered by the hot, X-ray emitting corona, should also account for the lack of observed relation between \gamx\ and \mbh\ (and for that matter, with \Lx; see again Fig.~\ref{fig:gam_tot_other_props}).

\newpage
\section{Conclusions}
\label{sec:conclusions}

We presented a detailed analysis of the links between the hard X-ray photon index, \gamx, and the (normalized) accretion rate, \lledd, for a large sample of hard X-ray selected, low-redshift AGN, as part of the BASS project.
Our analysis was motivated by several earlier studies that identified significant, positive correlations between \gamx\ and \lledd, over a broad range of redshifts.
The low-redshift BASS sample allowed us to study these relations over a wide range of \Lagn, \mbh, and \lledd. 
The high quality and broad spectral coverage of the BASS data -- unprecedented among studies that address the $\gamx-\lledd$ relation -- allowed us to examine, for the first time in this context,
the role of alternative determinations of the key quantities, and of the different methods used to derive them.
Our main conclusions are as follows:

\begin{enumerate}

\item
Despite a significant amount of scatter, we find a weak (but statistically significant) correlation between \gamx\ and \lledd\ among our primary sample of \Ngood\ AGN. 
This correlation is robust to the choice of \gamx. 

\item 
The best-fitting $\gamx-\lledd$ relations we obtain have flatter slopes than those reported by previous studies.
Moreover, these best-fitting relations fail to reduce the scatter in the $\gamx-\lledd$ plane.

\item
We find either no, or weak evidence for a $\gamx-\lledd$ correlation when considering, separately, the subsets of AGN that differ in the method used to derive \mbh\ (and therefore, \lledd). 
In particular, we find no correlation for the subset of AGN with the most reliable, ``direct'' mass estimates.

\item
We find no statistically significant correlations between \gamx\ and either the \lledd\ estimates based on \Lbat, nor with \Lhard, \mbh, or the width of the broad Balmer emission lines.

\item
A $\gamx-\lledd$ correlation that is consistent with those reported in previous studies \emph{does} emerge, for a subset of broad-line AGN, when adopting a simplified, power-law only spectral model fit to the lower-energy X-ray data, and only when coupled with \lledd\ determinations that are based on the \emph{optical} continuum emission.

\item
We caution that the prospects of using the $\gamx-\lledd$ relation for deriving distributions of \lledd\ (and indeed \mbh) from deep X-ray surveys are limited due to the large scatter in the $\gamx-\lledd$ plane, the weakness of the correlations we find, and their dependence on specific methodological choices (i.e., bolometric corrections and X-ray energy ranges).

\end{enumerate}

\noindent
Our analysis clearly demonstrates the complexity of the $\gamx-\lledd$ plane, even for a uniformly selected sample of nearby AGN, with a rich collection of multi-wavelength data, and a careful, elaborate spectral analysis.
It appears that the previously reported strong relations between \gamx\ and \lledd\ may be, at least partially, driven by methodological choices (i.e., \Lbol\ prescriptions) and/or limited spectral coverage and modelling in the X-ray regime.
Our results hint that an underlying physical mechanism that links the shape of the X-ray SED with \lledd\ may indeed be at work, but is not yet well understood.

The existence and robustness of the $\gamx-\lledd$ relation may be re-evaluated with yet larger, unbiased samples of hard X-ray selected AGN, provided by ongoing surveys using the \swift\ and {\it NuSTAR} missions. 
In particular, the {\it NuSTAR} mission is providing high sensitivity and high spatial resolution hard X-ray data for hundreds of AGN \cite[][]{Civano2015_NuSTAR_COSMOS,Lansbury2017_NuSTAR_serendip_40m}, reaching lower luminosities
and/or higher redshifts than previous hard X-ray studies.

\bigskip
\section*{Acknowledgements}

We thank M.\ Brightman, O.\ Shemmer, and R.\ B\"{a}r for their constructive comments, which helped us to improve the manuscript.
This work made use of the MATLAB package for astronomy and astrophysics \cite[][]{Ofek2014_matlab}.
This research has made use of the NASA/IPAC Extragalactic Database (NED),
which is operated by the Jet Propulsion Laboratory, California Institute of Technology, under contract with the National Aeronautics and Space Administration.
This research has made use of NASA's Astrophysics Data System.

CR acknowledges financial support from the China-CONICYT fund, FONDECYT 1141218 and Basal-CATA PFB-06/2007. 
MK acknowledges support from the Swiss National Science Foundation (SNSF) through the Ambizione fellowship grant PZ00P2\_154799/1.
KS and KO acknowledge support from the Swiss National Science Foundation (SNSF) through grants 200021\_157021, PP00P2\_138979, and PP00P2\_166159. 
ET acknowledges support from CONICYT-Chile grants Basal-CATA PFB-06/2007 and FONDECYT Regular 1160999.
The work of DS was carried out at the Jet Propulsion Laboratory,
California Institute of Technology, under a contract with NASA.
%

\newpage


\bibliographystyle{mnras}





\appendix
\section{Results of additional correlation analysis}
\label{app:sanity_checks}

As noted in Section~\ref{subsec:res_sanity_checks}, we have performed a series of correlation tests for different subsets of AGN in order to verify that our main results are not driven by the particular choices made through the sample definition and spectral analysis parts of our work.
Table~\ref{tab:corr_pars_minor} presents the results of the correlation hypothesis tests for these subsets, which include:

\begin{itemize}
\item \underline{AGN at $0.05<z<0.5$} -- a subset where the effects of spectroscopic aperture (relevant for \mbh, and there \lledd,  determination) are minimal.
For this sample we find no statistically significant correlation, neither when considering all \mbh\ subsets (84 sources), nor when considering each of these subsets separately (see $P$-values in Table~\ref{tab:corr_pars_minor}).

\item
\underline{AGN with SDSS-based optical spectroscopy} -- a subset where the (relative and absolute) flux calibration is optimal, and where aperture effects are small and well-understood.
For this sample of 47 sources we find a result similar to the general one: only the entire sample results in a significant correlation, while the relevant main \mbh\ subset (i.e., single-epoch estimates based on SDSS spectra) does not show a correlation.

\item
\underline{AGN with $0.01<\lledd<1$} -- a subset dominated by broad-line sources, which could in principle capture an underlying, positive $\gamx-\lledd$ correlation even if this relation flattens (or becomes an anti-correlation) at very low or very high \lledd.
For this sample of 195 sources (mostly broad-line AGN) we find a result similar to the general one: the entire sample results in a statistically significant, but weak correlation; the three \mbh\ subsets (within the $\lledd>0.01$ sample) do not show a correlation.

\item
\underline{AGN with $\log(\NH/\cmii)<23$} -- a subset where the effects of Compton scattering on the X-ray spectral decomposition are minimal.\\
For this sample of 188 sources we find, again, a result similar to the general one: only the entire sample results in a significant correlation, while the two main \mbh\ subsets (within the $\log(\NH/\cmii)<23$ sample) do not show a correlation.
For this subset, we also tested correlations involving \gamsoft\ and \gamnec, which are expected to be most sensitive to a Compton scattering component.
Indeed, we find that for this subset the correlations involving \gamsoft\ and \gamnec\ are somewhat stronger that those found with \gamtot. 
However, the qualitative outcome remains identical.

\item
\underline{AGN without warm absorbers} -- the presence of significant ionized absorption in the X-ray spectrum might lead to deviation of \gamx\ from the intrinsic value. For this sample of 184 sources we again find a result consistent with the general one: the entire sample results in a significant correlation, while the two main \mbh\ subsets do not show a correlation.

\end{itemize}

\begin{table*}
\centering
\caption{BASS $\gamx-\lledd$ correlations: significance tests for minor subsets}
\label{tab:corr_pars_minor}
\begin{tabular}{lclcrr} 
\hline
\Lbol\ & $\Gamma$ & sub-sample & $N$ &  $P-{\rm value}$ & $r_{\rm s}$ \\ 
tracer & ~~~      & ~~~        &     &  [\%]            & ~~~         \\
\hline
$C\cdot\Lhard$ & \gamtot   & $0.05<z<0.5$ ~- all                     &  84 &  8.1\%  & \ldots \\ 
~~~~~~~~~~~~~  & ~~~~~~~~  & ~~~~~~~~~~~~~~~~~~ - direct \mbh        &   9 & 61.3\%  & \ldots \\
~~~~~~~~~~~~~  & ~~~~~~~~  & ~~~~~~~~~~~~~~~~~~ - single-epoch \mbh\ &  64 &  9.9\%  & \ldots \\
~~~~~~~~~~~~~  & ~~~~~~~~  & ~~~~~~~~~~~~~~~~~~ - \sigs-based  \mbh\ &  11 & 32.7\%  & \ldots \\
\hline
$C\cdot\Lhard$ & \gamtot   & SDSS spec. ~- all                    &  47 &  1.70\% & \ldots \\ 
~~~~~~~~~~~~~  & ~~~~~~~~  & ~~~~~~~~~~~~~~~ - single-epoch \mbh\ &  33 & 54.7\%  & \ldots \\
~~~~~~~~~~~~~  & ~~~~~~~~  & ~~~~~~~~~~~~~~~ - \sigs-based  \mbh\ &  14 & 10.4\%  & \ldots \\
\hline
$C\cdot\Lhard$ & \gamtot   & $0.01<\lledd<1$ ~- all                   & 195 & $\mathbf{4.1\times10^{-2}\%}$ & 0.251 \\ 
~~~~~~~~~~~~~  & ~~~~~~~~  & ~~~~~~~~~~~~~~~~ - direct \mbh        &  29 & 63.2\%  & \ldots \\
~~~~~~~~~~~~~  & ~~~~~~~~  & ~~~~~~~~~~~~~~~~ - single-epoch \mbh\ & 140 &  0.83\% & \ldots \\
~~~~~~~~~~~~~  & ~~~~~~~~  & ~~~~~~~~~~~~~~~~ - \sigs-based  \mbh\ &  26 & 18.7\%  & \ldots \\
\hline
$C\cdot\Lhard$ & \gamtot   & $\log\NH<23$ ~- all                   & 188 & $\mathbf{1.7\times10^{-3}\%}$ & 0.308 \\ 
~~~~~~~~~~~~~  & ~~~~~~~~  & ~~~~~~~~~~~~~~~~ - direct \mbh        &  26 & 65.5\%  & \ldots \\
~~~~~~~~~~~~~  & ~~~~~~~~  & ~~~~~~~~~~~~~~~~ - single-epoch \mbh\ & 137 &  0.25\% & \ldots \\
~~~~~~~~~~~~~  & ~~~~~~~~  & ~~~~~~~~~~~~~~~~ - \sigs-based  \mbh\ &  25 & 39.5\%  & \ldots \\
\hline
$C\cdot\Lhard$ & \gamsoft  & $\log\NH<23$ ~- all                   & 188 & $\mathbf{1.6\times10^{-5}\%}$ & 0.371 \\ 
~~~~~~~~~~~~~  & ~~~~~~~~  & ~~~~~~~~~~~~~~~~ - direct \mbh        &  26 & 91.4\%  & \ldots \\
~~~~~~~~~~~~~  & ~~~~~~~~  & ~~~~~~~~~~~~~~~~ - single-epoch \mbh\ & 137 & $\mathbf{0.067\%}$ & 0.287 \\
~~~~~~~~~~~~~  & ~~~~~~~~  & ~~~~~~~~~~~~~~~~ - \sigs-based  \mbh\ &  25 & 39.5\%  & \ldots \\
\hline
$C\cdot\Lhard$ & \gamnec   & $\log\NH<23$ ~- all                   & 188 & $\mathbf{5.3\times10^{-6}\%}$ & 0.384 \\ 
~~~~~~~~~~~~~  & ~~~~~~~~  & ~~~~~~~~~~~~~~~~ - direct \mbh        &  26 & 93.8\%  & \ldots \\
~~~~~~~~~~~~~  & ~~~~~~~~  & ~~~~~~~~~~~~~~~~ - single-epoch \mbh\ & 137 & $\mathbf{0.0058\%}$ & 0.337 \\
~~~~~~~~~~~~~  & ~~~~~~~~  & ~~~~~~~~~~~~~~~~ - \sigs-based  \mbh\ &  25 & 88.0\%  & \ldots \\
\hline
$C\cdot\Lhard$ & \gamtot   & no warm absorbers - all               & 184 & $\mathbf{3.4\times10^{-3}\%}$ & 0.301 \\ 
~~~~~~~~~~~~~  & ~~~~~~~~  & ~~~~~~~~~~~~~~~~ - direct \mbh        &  20 & 69.1\%  & \ldots \\
~~~~~~~~~~~~~  & ~~~~~~~~  & ~~~~~~~~~~~~~~~~ - single-epoch \mbh\ & 115 & 0.65\%  & \ldots \\ 
~~~~~~~~~~~~~  & ~~~~~~~~  & ~~~~~~~~~~~~~~~~ - \sigs-based  \mbh\ &  49 & 39.2\%  & \ldots \\
\hline
\end{tabular}
\end{table*}


\bsp	
\label{lastpage}
\end{document}
